\documentclass[times]{elsarticle}

\usepackage{amssymb}

\usepackage[figuresright]{rotating}

\usepackage{url}
\usepackage{amsfonts}		
\usepackage{booktabs}  
\usepackage{caption}	
\usepackage{makeidx}         
\usepackage{graphicx}
\usepackage{subfig}	        
\usepackage{multicol}        
\usepackage[bottom]{footmisc}
\usepackage{epstopdf}
\usepackage{floatrow}		
\newfloatcommand{capbtabbox}{table}[][\FBwidth]
\usepackage{multirow}
\usepackage{blindtext}
\makeindex             
\usepackage{comment}
\includecomment{comment}
\makeatother
\usepackage{amsmath}
\allowdisplaybreaks 
\usepackage{array}
\usepackage{float}
\usepackage{longtable}
\usepackage{lscape}
\usepackage{algorithm}
\usepackage{algorithmic}
\usepackage{mathtools}
\usepackage{calc}  
\usepackage{enumitem}
\setlist[description]{leftmargin=\parindent,labelindent=\parindent}
\usepackage{graphicx} 
\usepackage{nomencl}
\makenomenclature 

\usepackage[numbers]{natbib}
\usepackage{amsthm}
\newtheorem{theorem}{Theorem}
\newtheorem{theorem1}{Special Theorem}
\newtheorem{definition}[theorem1]{Definition}
\usepackage[table]{xcolor}
\definecolor{Gray}{gray}{0.9}

\usepackage{color}
\usepackage{placeins} 

\long\def\AJP#1{{\bf[[AJP: #1]]}}

\long\def\ignore#1{}
\long\def\AJP#1{{\it\small\color{red}{[[AJP: #1]]}}}

\begin{document}

\begin{frontmatter}




\title{Two-Stage Stochastic International Portfolio Optimisation under Regular-Vine-Copula-Based Scenarios}


\author{Nonthachote Chatsanga}
\ead{psxnc2@nottingham.ac.uk}

\author{Andrew J. Parkes}
\ead{Andrew.Parkes@nottingham.ac.uk}

\address{The Automated Scheduling, Optimisation and Planning (ASAP) Group, School of Computer Science, The University of Nottingham, Nottingham NG8 1BB, UK}

\begin{abstract}

An international portfolio allows simultaneous investment in both domestic and foreign markets.  It hence has the potential for improved performance by exploiting a wider range of returns, and diversification benefits, than portfolios investing in just one market. However, to obtain the most efficient portfolios (along with the usual management of assets) the risks from currency fluctuations need good management, such as by using appropriate hedging. In this paper, we present a two-stage stochastic international portfolio optimisation model to find an optimal allocation for the combination of both assets and currency hedging positions. Our optimisation model allows a ``currency overlay'', or a deviation of currency exposure from asset exposure, to provide flexibility in hedging against, or in speculation using, currency exposure. The transaction costs associated with both trading and hedging are also included. 

To model the realistic dependence structure of the multivariate return distributions, a new scenario generation method, employing a regular-vine copula is developed. The use of vine copulas allows a better representation of the characteristics of returns, specifically, their non-normality and asymmetric dependencies. It hence improves the representation of the uncertainty underlying decisions needed for international portfolio optimisation problems. Efficient portfolios optimised with scenarios generated from the new vine-copula method are compared with the portfolios from a standard scenario generation method. Experimental results show that the proposed method, using realistic non-normal uncertainty, produces portfolios that give better risk-return reward than those from a standard scenario generation approach, using normal distributions.  The difference in risk-return compensation is largest when the portfolios are constrained to require higher returns. The paper shows that it can be important to model the non-normality in uncertainty, and not just assume normal distributions.

\end{abstract} 


\end{frontmatter}


\section{Introduction}

The seminal work of Markowitz \cite{Markowitz1952} formulated the portfolio optimisation problem by minimising risk, the variance of the returns, at a given level of mean or expected return. It is hence one method to handle optimisation under uncertainty.  A somewhat different standard method to cope with uncertainty in the input data is Stochastic programming \cite{BirgeLouveaux2011}, \cite{KaliWallace1994}, which can also be applied to an optimisation problems. Specifically, rather than representing input as a point (single) estimation, many possible estimations are represented, with each choice being called a scenario. Commonly, a discrete empirical probability distribution is employed to approximate a list of scenarios and so the corresponding probabilities \cite{Kaut2011}. The process to transform a given probability distribution to a set of scenarios is known as scenario generation; for a good presentation of such modelling and applications see Wallace and Ziemba \cite{Wallace2005}.

Typically, scenario generation methods characterise probability distributions by the first to the fourth statistical moments; that is, mean, variance (or standard deviation), skewness and kurtosis (See Larson \cite{Larson:1982nx} for more details). The relationship between different (marginal) distributions is generally described by a correlation matrix (see H{\o}yland et al. \cite{Hoyland2003} and Kaut et al. \cite{Kaut2007} for examples).  Although, these often suffices to capture the shape of marginal distributions, there still exist limitations in describing the relationship among distributions by such moments and correlations.  In particular, correlation (specifically the correlation coefficient, or Pearson's correlation \cite{Pearson1895}), captures the linear dependency between any two random variables, and does cannot capture non-linear dependencies, and so tends to give a misleading measurement when the measured data contain outliers or are highly-skewed. In such cases, an alternative treatment is to employ a rank correlation such as Kendall's tau \cite{Kendall1938}.  to measure a non-linear relationship. Nonetheless, for the ease of implementation, distributions are conventionally assumed normally-distributed when using correlations.

In reality, returns of most financial securities are non-Gaussian; and also they present an asymmetric dependence (see, for instance, Erb et al. \cite{Erb1994}, Longin and Solnik \cite{Longin2001}, Ang and Bekaert \cite{Ang2002a}, Ang and Chen \cite{Ang2002}, Campbell et al. \cite{Campbell2002}, Mitchell and Pulvino \cite{Mitchell2001} and Patton \cite{Patton2004}) in which returns are more strongly correlated in bear markets than in flat and bull markets. Thus, scenarios generated under normality and linear dependence assumptions do not reflect realistic events, and this has the potential to substantially affect the quality of solutions obtained by optimisation making such assumptions. As a consequence, methods are needed to extract appropriate information from the (historical) data, and then appropriately model, and exploit, the asymmetric non-normal dependencies among return distributions.  We will give methods based on copulas \cite{Sklar1959}, as they have become a standard method  in describing a generalised relationship among return distributions while retaining a good level of usability.   In particular, copulas have been used in scenario generation for stochastic programming in recent years, for example, in the works of Kaut \cite{Kaut2014}, Kaut and Wallace \cite{Kaut2011} and Sutiene and Pranevicius \cite{Sutiene2007}.

Sklar \cite{Sklar1959} describes a copula as a function that links a multidimensional distribution to its margins. The mathematical formulations of copulas are given in Sklar \cite{Sklar1996} and Nelsen \cite{Nelsen2007}. The major advantage of copulas is that they allow a separation of the marginal distributions and their dependency structures and hence these components can be modelled independently. This feature offers the flexibility to combine marginal distributions from different families within copulas; unlike a standard multivariate normal distribution where all marginal distributions are assumed Gaussian and with linear dependence. With copulas, the resulting scenarios generated can hence take into account non-normality such as heavy tails and asymmetric dependencies. Such improvements have the potential to help avoid risk underestimation in standard methods such as generating scenarios from using a multivariate normal distribution.
One of the contributions of this paper is an improvement of a scenario generation method. In the existing literature, \cite{Kaut2014} and \cite{Kaut2011} use empirical copulas whilst Sutiene and Pranevicius \cite{Sutiene2007} employ Gaussian and Student's t multivariate copulas in generating scenarios. The limitation of such empirical copulas is that the solutions could be unreliable when they are estimated from small samples. Also, applying multivariate copulas requires a dependence structure that can be described by one copula family; this lacks flexibility when used in high-dimensional modelling. To deal with such drawbacks, we exploit vine copulas \cite{Bedford2002} to model the dependence structure. In essence, a vine copula is a collection of bivariate copulas which can be represented by a nested set of trees that fulfil a certain set of conditions \cite{Bedford2002}. Vine copulas benefit from being able to use a wide range of bivariate copula families. A dependence structure of random variables will then be estimated pair-by-pair; a process which is more flexible than being characterised by a single copula family. The resulting scenarios generated with vine copulas should therefore have the potential to better represent dependencies of financial returns.

We will then exploit the vine-copula model of the real-world data, by a proposed two-stage stochastic international portfolio optimisation problem, and that is formulated with conditional value-at-risk (CVaR) as the risk measure. The formulation of an optimisation model is novel and it is another contribution of this paper.  The formulation also incorporates a currency overlay constructed with foreign exchange forwards, to allow currency exposure adjustments on a portfolio. Constraints associated to currency overlay and portfolio transactions are included in the optimisation model. Costs related to exchange rates hedging that affect risk and return of the portfolio are also taken into account.

The effects of a new scenario generation method and related constraints are empirically studied so as to investigate if the new approach produces portfolios that are more resilient to extreme events than a standard approach. More specifically, in our two-stage stochastic optimisation problem, portfolios are optimised with two types of scenarios. One from Regular-Vine-Copula-based scenarios as outlined earlier, hereafter referred to as an ``RVC portfolio''. Another is from a conventional method assuming that asset returns are normally distributed and relationship between asset returns is described by correlation; generally, a sampling is performed on a MultiVariate Normal distribution; portfolios optimised with these scenarios are henceforth referred to as an ``MVN portfolio''. Experiments are conducted accordingly to evaluate portfolio performances under the two types of scenarios.

The rest of this paper is organised as follows: Section \ref{sec:bckgnd} gives details on the background of copulas, regular-vine copulas, currency overlay, and CVaR.  Section \ref{sec:method} demonstrates an approach using a regular-vine copula to generate scenarios, a construction of currency overlay and a formulation of the optimisation model. Section \ref{sec:algo} describes an algorithm used in solving the optimisation problem. Section \ref{sec:res_disc} exhibits experiment results with analyses and section \ref{sec:concl} provides a conclusion of this study

\section{Background} \label{sec:bckgnd}

\subsection{Copulas}

Nelsen \cite{Nelsen2007} describes copulas as functions that join or ``couple'' multivariate distribution functions to their one-dimensional marginal distribution functions. The application in multivariate modelling was introduced by Sklar \cite{Sklar1959}, and demonstrated that a multivariate distribution can be decomposed into marginal distributions which are linked by copulas. Consequently, a given copula can produce various multivariate distributions by selecting different marginal distribution functions, and vice versa. Formally, Sklar's theorem is given by
\begin{theorem}[\textbf{Sklar's theorem}]
Let $F$ be an n-dimensional distribution function with margins $F_1, \ldots, F_n$. Then there exists a
unique copula $C$ such that
\begin{equation}
F(x_1, \ldots, x_n) = C\left(F_1(x_1), \ldots, F_n(x_n)\right)
\label{eq:Sklar_cdf}
\end{equation}
for all $x = (x_1, \ldots, x_n) \in \mathbb{R}^n$. If $F_1, \ldots, F_n$ are continuous, then $C$ is unique; otherwise, C is uniquely determined on $ \textrm{Ran}(F_1)\times\cdots\times \textrm{Ran}(F_n)$ which is the cartesian product of the ranges of the marginal cumulative distribution functions. Conversely, given a copula $C:[0,1]^n \rightarrow [0,1]$  and margins $F_i(x)$ then $C\left(F_1(x_1),\dots,F_n(x_n) \right)$ defines an n-dimensional cumulative distribution function.
\end{theorem}
A multivariate density function $f$ can be generated by differentiating (\ref{eq:Sklar_cdf}) using the chain rule as follows:
\begin{equation}
f(x_1, \ldots, x_n) = c\left(F_1(x_1), \ldots, F_n(x_n)\right)f_1(x_1) \ldots f_n(x_n).
\label{eq:Sklar_pdf}
\end{equation}

\subsection{Pair-Copula Construction and A Regular-Vine}

Theoretically, it is viable to construct higher-dimensional copulas with more than two variables, however, in practice, financial securities tend to have different dependence structure for each pair.  Characterising a dependence structure by a multivariate copula, assuming that all pairwise dependencies are alike, is thus inflexible and impractical.
Accordingly, we will model the dependence structure by a pair-copula construction (PCC). 
With a wide range of bivariate copulas available, the idea of PCC is to decompose a multivariate distribution into a product of bivariate copulas and marginal distributions. 

The structure of a pair-copula decomposition is can then be given by vines, a model of representing the construction steps and a dependence structure introduced by Bedford and Cook \cite{Bedford2002} (and that can also be well-represented graphically). A pair-copula construction starts with decomposing an $n$-dimensional joint density function $f$ as
\begin{equation}
f(x_1, \ldots, x_n) = f(x_1)f(x_2|x_1)f(x_3|x_2,x_1)\ldots f(x_n|x_{n-1}, \ldots, x_2, x_1).
\label{eq:multivar_decomposition}
\end{equation}
By the definition of conditional densities,
\begin{equation}
f(x_j|x_1,\ldots,x_{j-1}) = \frac{f(x_1,\ldots,x_{j-1},x_j)}{f(x_1,\ldots,x_{j-1})}.
\label{eq:conditional_density}
\end{equation}
According to Sklar's theorem, a joint density function $f$ can be expressed as a product of marginal density functions and their corresponding copula densities as in (\ref{eq:Sklar_pdf}). 
Combining (\ref{eq:Sklar_pdf}) and (\ref{eq:conditional_density}) allows the conditional densities to be expressed in terms of a pair-copula density and a marginal density. 
For instance,
\begin{equation}
f(x_1|x_2) = c_{12}\left(F_1(x_1), F_2(x_2)\right)f_2(x_2).
\label{eq:conditional_to_copula}
\end{equation}
The factorisation in (\ref{eq:conditional_to_copula}) is then a building block to higher-dimensional cases, which can be generalised to
\begin{equation}
f(x_i|\textbf{v}) = c_{x_i x_j|\textbf{v}_{-j}}\left(F(x_i|\textbf{v}_{-j}),F(x_j|\textbf{v}_{-j})\right)f(x_i|\textbf{v}_{-j})
\label{eq:conditionalPDF_generalised}
\end{equation}
for $i, j = 1,\ldots, n$ and $\textbf{v}$ denotes an arbitrary set of $x_1,\ldots, x_n$ including $x_j$ but not $x_i$ while $\textbf{v}_{-j}$ denotes all the elements from $\textbf{v}$ excluding $x_j$. Subsequently to (\ref{eq:conditionalPDF_generalised}), Joe \cite{Joe1996} shows that a conditional cumulative distribution function can be expressed in the following form:
\begin{equation}
F(x_i|\textbf{v}) = \frac{\partial C_{x_i x_j |\textbf{v}_{-j}}\left(F(x_i|\textbf{v}_{-j}), F(x_j|\textbf{v}_{-j})\right)}{\partial F(x_j|\textbf{v}_{-j})}
\label{eq:conditionalCDF_generalised}
\end{equation}
where $C_{x_i x_j |\textbf{v}_{-j}}$ is a bivariate copula distribution function to $F_{x_i x_j |\textbf{v}_{-j}}$. Applying (\ref{eq:conditionalPDF_generalised}) to all conditional densities in (\ref{eq:multivar_decomposition}) gives the following decomposition:
\begin{equation}
f(x_1, \ldots, x_n) = \prod\limits_{i=1}^n f(x_i)\prod\limits_{i=2}^n\prod\limits_{j=1}^{i-1}c_{ij|(j+1)\ldots(i-1)}\left( F(x_i|x_{j+1},\ldots,x_{i-1}),F(x_j|x_{j+1},\ldots,x_{i-1}) \right)
\label{eq:PCC_decomposition}
\end{equation}
and there can be numerous ways in which to decompose (\ref{eq:PCC_decomposition}).
The number of decompositions increases rapidly with the dimension of random variables. It is, therefore, necessary to employ a tool such as a regular-vine (R-Vine) for organising the large number of pair-copula constructions. Other variants of vine copulas, as special cases of an R-Vine, are a canonical-vine (C-Vine) and a drawable-vine (D-Vine). More details on vine-copulas are given in Aas et al. \cite{Aas2009}. A vine is given by a set of trees; random variables are nodes of a tree and edges between nodes represent copulas. In the first tree, all pairs are unconditioned to other variables. In the second tree, all pairs are conditioned on one other variable. In the third tree, all pairs are conditioned on two variables, and so on. The formal definition of regular-vine copulas is given in Bedford and Cooke \cite{Bedford2002} and Kurowicka and Cooke \cite{Kurowicka2006} as
\begin{definition}[\textbf{R-Vine}]
$\mathcal{V} = (T_1,\ldots, T_{n-1})$ is an R-Vine on $n$ elements if
\begin{enumerate}[label=(\roman*)]
\item $T_1$ is a tree with nodes $N_1 = \{1,\ldots, n\}$ and a set of edges denoted by $E_1$.
\item For $i = 2,\ldots,n-1$, $T_i$ is a tree with nodes $N_i = E_{i-1}$ and an edge set $E_i$.
\item For $i = 2,\ldots,n-1$ and $\{a,b\} \in E_i$ with $a = \{a_1, a_2\}$ and
$b = \{b_1, b_2\}$ it must hold that $\#(a \cap b) = 1$ (proximity condition).
\end{enumerate}
\end{definition}

Briefly speaking, regular-vines are a graph-based tool for specifying conditional bivariate constraints. An R-Vine of $n$-dimensions is a nested set of $n-1$ trees and $n(n-1)/2$ edges such that the nodes of tree $i + 1$ are the edges of tree $i$ and two nodes of tree $i + 1$ are connected by an edge only if they share a common node in tree $i$ (the proximity condition). 

\subsection{Conditional Value-at-Risk (CVaR)}

A selection of an appropriate risk measure is vital to portfolio optimisation problems. The Basel III regulatory framework formulates a risk measure as a percentile (generally $5^{th}$ percentile) of a loss distribution, i.e., a Value-at-Risk (VaR). It represents a maximum loss under specified probability (confidence level) over a certain time period. In order to define a Value-at-Risk and a Conditional Value-at-Risk (CVaR), we first need to define a loss function which represents negative returns of a portfolio, where the portfolio return is given by the summation of individual asset return $\xi_i$ weighted by an allocation $w_i$. 
\begin{equation}
f_L(w,\xi) = -\sum\limits_{i}w_i \xi_i. \nonumber
\end{equation}  
Then, a probability that a loss $f_L(w,\xi)$ does not exceed a threshold $\alpha$ is
\begin{equation}
\Phi(w,\alpha) = \int_{f_L(w,\xi)\leq \alpha}p(\xi)d\xi \nonumber
\end{equation}
where $p(\xi)$ is a joint density function of random returns and $\Phi(w,\alpha)$ is a cumulative
distribution loss function associated with $w$ which is continuous and non-decreasing with respect to $\alpha$. Formally, VaR with respect to the portfolio weights $w$ at a confidence level $\beta \in (0,1)$ is given by the smallest $\alpha$ such that the probability of the loss $f_L(w,\xi)$ exceeding $\alpha$ is at most $1-\beta$ as follows:
\begin{equation}
\text{VaR}_\beta(w) = \text{inf}\{\alpha: \Phi(\xi,\alpha) \geq \beta \}. \nonumber
\end{equation}
However, by its definition, VaR does not distinguish the extent of losses beyond a threshold. Besides,
the non-convex characteristic of VaR implies that minimising VaR does not guarantee a global minimum. In addition, Rockafellar and Uryasev \cite{Rockafellar2002} address the drawback of VaR that it is unstable and difficult to handle numerically when dealing with non-normal distributions. All these shortcomings disallow VaR from being an appropriate risk measure for portfolio optimisation problems.

\begin{figure}[H]
\begin{centering}
\includegraphics[width=3in]{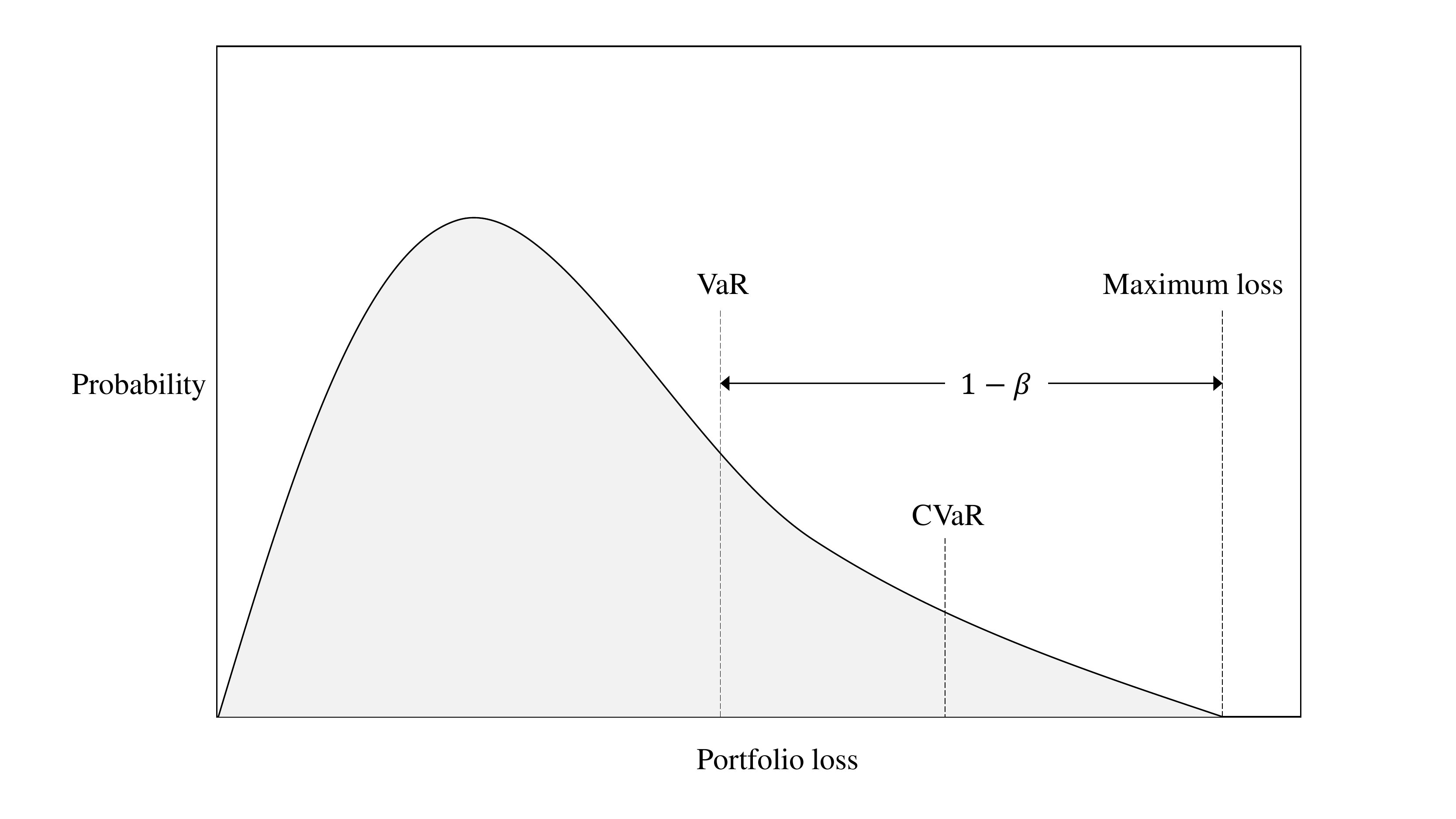} 
\par\end{centering}
\caption{Value-at-Risk (VaR) and Conditional Value-at-Risk (CVaR). CVaR represents the expected loss exceeding VaR.}
\label{CVaR}
\end{figure}

Artzner et al. \cite{Artzner1999} suggest the desirable properties of risk measures leading up to the notion of coherent risk measures. A Conditional Value-at-Risk (CVaR) which estimates the expected loss greater than VaR satisfies all the criteria for coherence. Formally, CVaR with respect to a portfolio allocation $w$ at a confidence level $\beta \in (0,1)$ is defined by
\begin{equation}
\text{CVaR}_\beta(w) = \frac{1}{1-\beta}\int_{f_L(w,\xi)\geq \text{VaR}_\beta(w)}f_L(w,\xi)p(\xi)d\xi.
\end{equation}
Note that the probability that $f_L(w,\xi)$ exceeding VaR accumulates to $1-\beta$. The relationship between VaR and CVaR is illustrated in Figure \ref{CVaR}. Rockafellar and Uryasev \cite{Rockafellar2000} introduce an auxiliary function to compute VaR and CVaR as follows:
\begin{align}
F_\beta(w,\alpha) &= \alpha+\frac{1}{1-\beta}\int_{f_L(w,\xi)\geq \alpha}\left(f_L(w,\xi)-\alpha\right)p(\xi)d\xi, \nonumber \\
&= \alpha+\frac{1}{1-\beta}\mathbb{E}[ \left( f_L(w,\xi)-\alpha \right)^+ ]
\label{eq:CVaR_auxeq}
\end{align}
where $\mathbb{E}[\cdot]$ is an expectation operator and $\left( f_L(w,\xi)-\alpha \right)^+ = \text{max}\left( f_L(w,\xi)-\alpha,0 \right)$. Practically, the true joint density $p(\xi)$ is often unknown and needed to be estimated. A discrete approximation of $p(\xi)$ are generally used to represent the joint density of portfolio returns. Accordingly, the corresponding approximation of $F_\alpha(w,\alpha)$ in (\ref{eq:CVaR_auxeq}) given scenarios $s = 1,\ldots,N$ is
\begin{equation}
\widetilde{F}_\beta(w,\alpha) = \alpha+\frac{1}{(1-\beta)N}\sum\limits_{s=1}^N\left( f_L(w,\xi_s)-\alpha \right)^+
\end{equation}
The function approximating CVaR value $\widetilde{F}_\beta(w,\alpha)$ can be transformed into a linear expression by replacing $\left( f_L(w,\xi_s)-l \right)^+$ with an artificial variable $e_s$ as
\begin{equation}
\widetilde{F}_\beta(w,\alpha) = \alpha + \frac{1}{(1-\beta)N}\sum\limits_{s=1}^N e_s 
\end{equation}
where
\begin{align}
e_s &\geq f_L(w,\xi_s)-\alpha, \\
e_s &\geq 0.
\end{align}
When applying CVaR in optimisation problems, it is shown in \cite{Rockafellar2000} that mean-CVaR and mean-variance portfolios produce the same efficient frontier if a loss function is normally distributed. However, a difference between the two approaches, can occur when the underlying distribution is non-normal and asymmetric, and so is important in this paper.

\subsection{Currency Overlay}

In this section, we give a background on a currency overlay, based on our previous work in Chatsanga and Parkes \cite{NCAP2016} where more details can be found.
 
When a portfolio invests in several countries with different currencies, a portfolio value is determined by two factors. One is from asset prices plus dividends or other interest-rate-bearing incomes and another one is from gain or loss of exchange rates. Investment in each country is thus portrayed as a composition of exposure in asset markets and exposure in exchange rates. Structuring a portfolio in this way provides flexibility in adjusting currency exposure arising from foreign currency positions. An alteration made on currency exposure is regarded as currency overlay which modifies the status-quo currency positions of an unhedged portfolio. 

A currency overlay \cite{Topaloglou2008} is defined as a deviation of currency exposure from asset exposure in a portfolio. Such a deviation is created by holding foreign exchange forward contracts (FX forwards) which are widely used in managing currency exposure \cite{FonsecaRustem2012} and are available in various currency pairs. Holding each forward pair increases exposure in one currency and decreases exposure in another currency. For instance, having a position of EURUSD forward at 1\% of a portfolio value increases EUR exposure by 1\% and reduces USD exposure by 1\%.

Since each forward contract involves only a pair of currencies, given that there are $C$ countries to invest, then there are 
$K = \binom{C}{2}$   
\AJP{Note: using the "binom" works without errors } 
different forward contracts in total. 
The forward position denoted by $f_{kj}$ represents how much additional exposure is added into or taken off from each currency by the $K$-th forward contract. 
Since the exposure from a pair of currencies are equal when being valued in a portfolio's base currency, then it is strictly required that $\underset{j}\sum f_{kj} = 0$ for all $k=1,...,K$. 

Denoting $\textbf{f}_{k}=(f_{k1}, \ldots, f_{kC})$ a vector of exposure from all possible forward contracts, the previous requirement implies that only two elements of $\textbf{f}_{k}$ represent the exposure in which one being equal to a negative value of another, while the rest of the elements only takes a value of zero. To avoid putting those requirements into constraints of an optimisation problem, we define a matrix $\textbf{F}$ with $f_{kj}$ representing its elements and $\textbf{f}_{k}$ representing its rows. The matrix $\textbf{F}$ is then constructed by:
\begin{equation}
\textbf{F}  \,\stackrel{\text{def}}{=}\, \textbf{T} \circ (\textbf{1}^{\textrm{T}} \otimes \textbf{q})
\label{Fmatrix}
\end{equation}
where $\circ$ is the Hadamard product operator, $\otimes$ is the Kronecker product operator, $\textbf{T}$ is a $K \times C$ combinatorial matrix of $\{-1,0,1\}$, $\textbf{1}$ is a $C \times 1$ column vector of ones and $\textbf{q} = (q_1, \ldots, q_K)$ is a $K \times 1$ column vector determining the size of exposure.

The combinatorial matrix $\textbf{T}$ of ternary variables is constructed by first specifying $\textbf{T}$ as a matrix of zeros of size $K \times C$, then we denote a set $\textbf{D}$ containing all combinatoric pairs from ${\binom{C}{2}}$. In each row of $\textbf{T}$, the first member of each pair in $\textbf{D}$ specifies which element to take the value of 1 and the second member of the pair specifies the element that take value of -1. The structure of an international portfolio with currency overlay is presented in Table \ref{table:overlay_structure_1}.

Suppose that a portfolio manager plans to invest in $A$ asset classes from $C$ countries, then the exposure of assets and currencies in a portfolio can be characterised as in Table \ref{table:overlay_structure_1} with the following notations:
\begin{description}[leftmargin=!,labelwidth=\widthof{\bfseries cccccc}]
\item[\textnormal{i}] Index of asset classes; $i=1, ..., A$
\item[\textnormal{j}] Index of countries, or synonymously currencies; $j=1, ..., C$
\item[\textnormal{k}] Index of forward contracts $k=1, ..., K$
\item[\textnormal{$\tilde{a}_{ij}$}] Exposure (allocation) to asset class $i$ of country $j$
\item[\textnormal{$f_{kj}$}] Forward position (allocation) of contract $k$ on country (currency) $j$
\end{description}

\begin{table}[H]
\resizebox{\linewidth}{!}{
\centering
{\small
\begin{tabular}{lllllll}
\toprule
 &  & Country 1 & $\cdots$ & Country j & $\cdots$ & Country C\tabularnewline
\midrule
Asset class 1 &  & $\tilde{a}_{11}$ & $\cdots$ & $\tilde{a}_{1j}$ & $\cdots$ & $\tilde{a}_{1C}$\tabularnewline
$\vdots$ &  & $\vdots$ & $\ddots$ & $\vdots$ & $\ddots$ & $\vdots$\tabularnewline
Asset class $i$ &  & $\tilde{a}_{i1}$ & $\cdots$ & $\tilde{a}_{ij}$ & $\cdots$ & $\tilde{a}_{iC}$\tabularnewline
$\vdots$ &  & $\vdots$ & $\ddots$ & $\vdots$ & $\ddots$ & $\vdots$\tabularnewline
Asset class $A$ &  & $\tilde{a}_{A1}$ & $\cdots$ & $\tilde{a}_{Aj}$ & $\cdots$ & $\tilde{a}_{AC}$\\[.2cm]
\rule{0pt}{2.5ex}Forward position 1 &  & $f_{11}$ & $\cdots$ & $f_{1j}$ & $\cdots$ & $f_{1C}$\tabularnewline
$\vdots$ &  & $\vdots$ & $\ddots$ & $\vdots$ & $\ddots$ & $\vdots$\tabularnewline
Forward position $k$ &  & $f_{k1}$ & $\cdots$ & $f_{kj}$ & $\cdots$ & $f_{kC}$\tabularnewline
$\vdots$ &  & $\vdots$ & $\ddots$ & $\vdots$ & $\ddots$ & $\vdots$\tabularnewline
Forward position $K$ &  & $f_{K1}$ & $\cdots$ & $f_{Kj}$ & $\cdots$ & $f_{KC}$\\[.2cm]
\rule{0pt}{2.5ex}Asset exposure &  & $\sum\limits_{i=1}^{A}\tilde{a}_{i1}$ & $\cdots$ & $\sum\limits_{i=1}^{A}\tilde{a}_{ij}$ & $\cdots$ & $\sum\limits_{i=1}^{A}\tilde{a}_{iC}$\\[.2cm]
Overlay position &  & $\sum\limits_{k=1}^{K}f_{k1}$ & $\cdots$ & $\sum\limits_{k=1}^{K}f_{kj}$ & $\cdots$ & $\sum\limits_{k=1}^{K}f_{kC}$\\[.2cm]
Currency exposure &  & $\sum\limits_{i=1}^{A}\tilde{a}_{i1}+\sum\limits _{k=1}^{K}f_{k1}$ & $\cdots$ & $\sum\limits_{i=1}^{A}\tilde{a}_{ij}+\sum\limits _{k=1}^{K}f_{kj}$ & $\cdots$ & $\sum\limits_{i=1}^{A}\tilde{a}_{iC}+\sum\limits _{k=1}^{K}f_{kC}$\\[.2cm]
Total overlay &  & \multicolumn{5}{l}{$\frac{1}{2}\sum\limits_{j=1}^{C}\left|\sum\limits _{k=1}^{K}f_{kj}\right|$}\tabularnewline
\bottomrule
\end{tabular} 
}
\caption{Structure of a multi-currency portfolio with a currency overlay.}
\label{table:overlay_structure_1}
}
\end{table}

\subsection{Cost of Carry of Foreign Exchange Forwards}

This section gives a background on costs associated with holding forward contracts in a portfolio as given in \cite{NCAP2016}, where more details can be found. In our portfolio, a currency overlay can be created from several foreign exchange forwards. Costs associated to holding a currency overlay is thus important in determining return and risk of a portfolio. Apart from transaction costs \cite{Bertsimas2008} occurring from buying or selling assets, holding a forward contract also incurs a cost from interest rate differential of corresponding currencies in a forward contract which is known as a ``cost of carry''. 

The relationship between the values of a spot rate and a forward rate is determined by a difference in interest rates earned on the respective currency pairs. The idea is that buying a forward contract is equivalent to buying an underlying asset now and paying a ``carry'' until the end of the contract.

According to \cite{NCAP2016}, a cost of carry from holding a foreign exchange forward contract can be computed by
\begin{equation}
\label{CoC}
\text{Cost of Carry} = i_{buy} - i_{sell}	
\end{equation}
where $i_{sell}$ is an interest rate of a country that one wants to sell the currency off so as to buy another currency and $i_{buy}$ is an interest rate of a country that one desires to buy.

Entering into forward contracts incurs a cost of carry which can be positive or negative depending on interest rate differential. Consider a portfolio that holds three foreign exchange forwards as given in Table~\ref{tab:CoC}. A cost of carry with respect to each forward contract depends on which currency to sell or buy, corresponding interest rates, and a position taken on the portfolio. For instance, selling JPY for USD at 1\% of the portfolio amounts the positive carry of $1\%\times2\%-1\%\times1\% = 0.01\%$ to the portfolio. Selling GBP for JPY, however, generates a negative carry of $-2\%\times4\%+2\%\times1\% = -0.06\%$ as a result of shifting exposure from the country with high interest rate to the country with lower interest rate. The total overlay position is 8\% of the portfolio bearing the positive carry of 0.13\% from the three forward contracts combined. This amount of carry is added to the total return of the portfolio.

\begin{table}[bthp]
  \centering
  \caption{Costs of carry associated with foreign exchange rate forward contracts. Numbers in bold indicate positions on a portfolio (in percentage). The total currency overlay position on each currency is calculated from the net forward positions on the respective currency. The cost of carry from holding each forward contract is the weighted sum of interest rates and forward positions with respect to currencies associated with the forward contract.}
    \begin{tabular}{rcccc}
    \toprule
          & \small{USD}   & \small{GBP}   & \small{JPY}   & {\small Cost of Carry} \\
    \midrule
    \small{interest rate (\%)} & {\small 2}   & {\small 4}   & {\small 1}   &  \\
    \rule{0pt}{2.5ex}{\small sell JPY, buy USD (\%)} & \textbf{\small 1}   &       & \textbf{\small -1}  & {\small 0.01} \\
    {\small sell USD, buy GBP (\%)} & \textbf{\small -9}  & \textbf{\small 9}   &       & {\small 0.18} \\
    {\small sell GBP, buy JPY (\%)} &       & \textbf{\small -2}  & \textbf{\small 2}   & {\small -0.06} \\
    \rule{0pt}{2.5ex}{\small overlay (\%)} & \textbf{\small -8}  & \textbf{\small 7}   & \textbf{\small 1}   & {\small 0.13} \\
    \bottomrule
    \end{tabular}
  \label{tab:CoC}
\end{table}

From Table \ref{tab:CoC}, the net cost of carry is in fact a product of interest rates and overlay positions. For an investment in any country $j$, a total return contributes to a portfolio is
\begin{equation}
\label{ret_j}
r_j = \tilde{a}_{j}r^a_j + c_{j}r^c_j + v_{j}i_j
\end{equation}
where $r_j$ is a total return from investment in a country $j$; $\tilde{a}_j$, $c_j$ and $v_j$ are respectively asset exposure, currency exposure and an overlay position on country $j$; $r^a_j$, $r^c_j$ and $i_j$ are respectively expected asset return, expected currency return and expected interest rate of a country $j$.

Since an overlay position is defined as a difference in currency and asset exposure, equation (\ref{ret_j}) can be equivalently expressed as
\begin{align}
r_j &= \tilde{a}_{j}r^a_j + c_{j}r^c_j + (c_{j}-\tilde{a}_{j})i_j \nonumber \\
    &= \tilde{a}_{j}(r^a_j - i_j) + c_{j}(r^c_j + i_j).
\label{adjusted_return}
\end{align}
We define $r^a_j - i_j$ and $r^c_j + i_j$ as an adjusted return of asset and an adjusted return of currency, respectively. Equation (\ref{adjusted_return}) demonstrates that a portfolio total return (return from assets, currencies and costs of carry of foreign exchange forwards) is equal to a product of adjusted returns, asset exposure and currency exposure. This implies that the expression of overlay positions is not explicitly required in order to calculate total returns of a portfolio. In addition, if a portfolio holds no forward contract, the interest rate terms in equation (\ref{adjusted_return}) will be cancelled out, showing that the formulation in equation (\ref{adjusted_return}) generalises total return calculation of an international portfolio.

Similarly to asset and currency returns, interest rates are not constant over time, the volatility of interest rates is thus needed to be included in calculating portfolio risk. In accordance with return calculation of international portfolios, we apply equation (\ref{adjusted_return}) to adjust return time series before generating scenarios. 
\section{Methodology} \label{sec:method}

\subsection{Scenario Generation with Regular-Vine Copula Dependence Structure}\label{RVC_scenario_generation}

A procedure to generate scenarios for our optimisation problem in this paper is described below. An example of generating scenarios with an R-Vine copula on three variables using the following method is demonstrated in \ref{apdx:ScenarioGeneration}.

\begin{enumerate}
\item \textit{Modelling marginal distributions} - Given empirical data in the form of time series of financial returns, we separately fit an invertible empirical distribution to each time series and estimate a marginal probability distribution function (PDF). We employ a kernel density estimation (KDE) to estimate an empirical PDF of a return time series. To approximate a probability density function $f$ of a random variable $\Xi_i$ for $i=1,\ldots,n$ assuming that we have $m$ independent observations $\xi_{i1},\ldots,\xi_{im}$ on each random variable, a kernel density estimator for the estimation of the density value at point $x$ is defined as
\begin{equation}
{f}(\xi_i)= \frac{1}{mh}\sum\limits_{j=1}^{m} K\left(\frac{\xi_{ij}-\xi_i}{h}\right)
\end{equation}
where $K$ denotes a so-called kernel function, and $h$ is a bandwidth. In our study, we select the Epanechnikov kernel as a kernel function and choose an optimal bandwidth according to Silverman's rule of thumb, \cite{Silverman1986}.

An empirical cumulative distribution function (CDF) of each return series can be subsequently produced from an estimated PDF as follows:
\begin{equation}
F(\xi_i) = \sum\limits_{\xi_{ij} \leq \xi_i} {f}(\xi_{ij}).
\end{equation}
The resulting CDF is uniform on the interval $\left[0,1\right]$ and is an input argument to a copula function. In what follows, we denote a CDF of a random variable $i$ by $u_i$. 

\item \textit{Estimating a regular-vine copula} - To fit an R-Vine copula to a given dataset, Dissmann et al. \cite{Dissmann2012} outline the procedure as follows:
\begin{enumerate}[label=(\alph*)]
\item Selection of the R-Vine structure, i.e. selecting which unconditioned and conditioned pairs to use for the PCC. \label{task1}
\item Fitting a pair-copula family to each pair selected in \ref{task1}.
\item Estimation of the corresponding parameters for each copula.
\end{enumerate}
In our study, the sequential method developed by Dissmann \cite{Dissmann2010} which fits an R-Vine tree-by-tree is employed to estimate the R-Vine copula. The bivariate copula families involving in the model estimation include elliptical (Gaussian and Student-t) as well as Archimedean (Clayton, Gumbel, Frank, Joe, BB1, BB6, BB7 and BB8) copulas to cover a broad range of possible dependence structures. For the Archimedean copula families, rotated versions are also included so as to cover negative dependencies. Thorough details of the bivariate copula families are provided in \cite{Joe1997} and \cite{Nelsen2007}.

In our study, an R-Vine model is estimated using the \texttt{VineCopula} package on R, \cite{SchepsmeierStoeberBrechmannEtAl2012}. The results of this step provide an exact combination of bivariate copulas and conditional bivariate copulas (with respect to an R-Vine structure) that is most most appropriate to the given data.

\item \textit{Sampling from a regular-vine density} - We follow the R-Vine sampling method from \cite{Dissmann2012}. This step starts with sampling $u_1,\ldots,u_n$ which are independent and uniform on $\left[0,1\right]$ then set 
\begin{align}
\begin{split}
\xi_1 &= u_1, \\
\xi_2 &= F^{-1}_{2|1}(u_2|\xi_1), \\
\xi_3 &= F^{-1}_{3|12}(u_3|\xi_1,\xi_2), \\
    &\vdots \\
\xi_n &= F^{-1}_{n|12...n-1}(u_n|\xi_1,\ldots,\xi_{n-1})   
\end{split}
\label{eq:RVine_sampling} 
\end{align}
where $F^{-1}_{j|12...j-1}(u_n|\xi_1,\ldots,\xi_{j-1})$ for $j=1,\ldots,n$ is the inverse of (\ref{eq:conditionalCDF_generalised}). (\ref{eq:RVine_sampling}) produces dependent $\xi_1,\ldots,\xi_n$ which is equivalent to one scenario for random variables $\Xi_1,\ldots,\Xi_n$. To generate $N$ scenarios, the random sampling of $u_1,\ldots,u_n$ is repeated $N$ times.
\end{enumerate}

\subsection{Two-Stage Stochastic International Portfolio Optimisation Model}

\subsubsection{Notations}
The notations used in formulating the optimisation model are categorised by their types and presented in
separate tables. Table \ref{tab:notations_model} provides the notations for parameters in the optimisation model. The values are set arbitrarily and can be adjusted to individual preferences. Table \ref{tab:notations_transaction} shows the notations of variables related to transaction costs. The variable costs are retrieved from market data while the fixed costs are set arbitrarily. Table \ref{tab:notations_auxvar} displays auxiliary variables that facilitate calculations of the fitness function and constraints. Table \ref{tab:notations_decvar} exhibits the associated decision variables.

Note that in Table \ref{tab:notations_auxvar}, $a_{ij}$ denotes final units of an asset class $i$ in country $j$ which differs from its corresponding exposure $\tilde{a}_{ij}$ defined earlier in Table \ref{table:overlay_structure_1}. To calculate the exposure $\tilde{a}_{ij}$, let $P_{ij}^0$ be a market price of $a_{ij}$, then $\tilde{a}_{ij} = a_{ij}P_{ij}^0$.

\begin{table}[H]
  \centering
  \caption{Notations for the optimisation model parameters.}
  {\small
    \begin{tabular}{p{1.5cm}p{1cm}p{8.5cm}}
    \toprule
    Notation & Range & Description \\
	\midrule  
    $A$ & & A set of assets. \\
    $C$ & & A set of currencies. \\
    $G$ & & A set of foreign exchange forwards. \\
    $N_r$ & & A set of recourse nodes, each node corresponds to each recourse portfolio.\\    
    $i$ & $\mathbb{N}$ & An index of asset classes, $i \in A$. \\
    $j$ & $\mathbb{N}$ & An index of currencies, $j \in C$. \\
    $k$ & $\mathbb{N}$ & An index of foreign exchange forwards, $k \in G$. \\
    $\mu$ & $\mathbb{R^+}$ & A return target of a portfolio. \\
    $\beta$ & $[0,1]$ & The quantile for CVaR. \\ [3pt]   
    $a_{ij}^{min}$, $a_{ij}^{max}$ & $\mathbb{R^+}$ & Minimum and maximum holding positions of an asset $a_{ij}$, respectively. \\
    $c_j^{min}$, $c_j^{max}$ & $\mathbb{R^+}$ & Minimum and maximum currency exposures of currency $j$ allowed on a portfolio. \\
    $t_{ij}^{min}$, $t_{k}^{max}$ & $\mathbb{R^+}$ & A minimum trading size of an asset $a_{ij}$ and a forward pair $q_k$. \\ [3pt]
    $V_u$ & $[0,1]$ & A total currency overlay limit on a portfolio. \\ [3pt]
    $K_C$, $K_G$ & $\mathbb{N}$ & The number of countries allowed to invest and the number of forward pairs allowed to hold in a portfolio (cardinality). \\   
    $M$ & $[0,1]$ & A percentage of cash margin required to hold a forward contract. \\
    $B$ & $\mathbb{R^+}$ & An arbitrary big constant. \\
    $h^0$ & $\mathbb{R^+}$ & An initial cash amount in a portfolio. \\
    $a_{ij}^0$ & $\mathbb{R^+}$ & An initial position of an asset class $i$ in currency $j$. \\[3pt]
    $q_{k}^0$ & $\mathbb{R^+}$ & An initial position of a forward pair $k$. \\[3pt] 
    $W^{0}$ & $\mathbb{R^+}$ & An initial wealth of a portfolio. \\
    $p^{r}$ & $[0,1]$ & A probability of a recourse node $r$ in the second stage. \\
    \bottomrule
    \end{tabular}%
  \label{tab:notations_model}%
  }
\end{table}%

\begin{table}[H]
  \centering
  \caption{Notations for the transaction costs.}
  {\small
    \begin{tabular}{p{1.5cm}p{1cm}p{8.5cm}}
    \toprule
    Notation & Range & Description \\
	\midrule  
    $\pi_{ij}$, $\pi_{k}$ & $\mathbb{R^+}$ & Fixed costs for buying an asset $a_{ij}$ and a forward pair $q_{k}$, respectively. \\
    $\psi_{ij}$, $\psi_{k}$ & $\mathbb{R^+}$ & Fixed costs for selling an asset $a_{ij}$ and a forward pair $q_{k}$, respectively. \\
    $\rho_{ij}$, $\rho_{k}$ & $\mathbb{R^+}$ & Variable costs for buying an asset $a_{ij}$ and a forward pair $q_{k}$, respectively. \\
    $\lambda_{ij}$, $\lambda_{k}$ & $\mathbb{R^+}$ & Variable costs for selling an asset $a_{ij}$ and a forward pair $q_{k}$, respectively. \\     
    $P_{ij}^{0}$, $P_{k}^{0}$ & $\mathbb{R^+}$ & Prices (in base currency) per unit of an asset $a_{ij}$ and a forward pair $q_{k}$ in the first stage, respectively. \\
    \bottomrule
    \end{tabular}%
  \label{tab:notations_transaction} %
  }
\end{table}%

\begin{table}[H]
  \centering
  \caption{Notations for the auxiliary variables of the optimisation model.}
  {\small
    \begin{tabular}{p{1.5cm}p{1cm}p{8.5cm}}
    \toprule
    Notation & Range & Description \\
	\midrule  
    $\textbf{F}$, $\textbf{F}^{r}$ & & A matrix specifying currency exposure adjustment corresponding to market values of forward positions held in the first stage and at a recourse node $r$, respectively. \\
    $\textbf{T}$ & & A ternary matrix containing 0,1 and -1. \\
    $\textbf{q}$, $\textbf{q}^{r}$ & & A vector containing market values of forward positions in the first stage and at a recourse node $r$, respectively. \\    
    $f_{kj}$, $f_{kj}^{r}$ & $\mathbb{R}$ & A matrix element of $\textbf{F}$ in the first stage and at a recourse node $r$, respectively. \\
    $W^{r}$ & $\mathbb{R}$ & Wealth of a portfolio at a recourse node $r$. \\
    $c_{j}$, $c_{j}^{r}$ & $\mathbb{R}$ & Currency exposure of currency $j$ in the first stage and at a recourse node $r$, respectively. \\
    $e^{r}$ & $\mathbb{R}$ & Expected shortfall (expected loss exceeding VaR) at a recourse node $r$. \\
    $\alpha$ & $\mathbb{R}$ & VaR value. \\
    $a_{ij}$, $q_{k}$ & $\mathbb{R}$ & Final units of an asset $a_{ij}$ and a forward pair $q_{k}$ in the first stage. \\
    $a_{ij}^{r}$, $q_{k}^{r}$ & $\mathbb{R}$ & Final units of an asset $a_{ij}$ and a forward pair $q_{k}$ at a recourse node $r$, respectively. \\
    $a_{ij}^{csh}$, $a_{ij}^{csh,r}$ & $\mathbb{R}$ & An amount of cash reserved for margin requirement of forwards (considered as an asset class denominated in base currency) in the first stage and at a recourse node $r$, respectively. \\   
    \bottomrule
    \end{tabular}%
  \label{tab:notations_auxvar}%
  }
\end{table}%

\begin{table}[H]
  \centering
  \caption{Notations for the decision variables of the optimisation model.}
  {\small
    \begin{tabular}{p{1.5cm}p{1cm}p{8.5cm}}
    \toprule
    Notation & Range & Description \\
	\midrule  
    $b_{ij}$, $b_{k}$ & $\mathbb{R^+}$ & The number of units of an asset $a_{ij}$ and the number of units of a forward pair $q_{k}$ bought in the first stage, respectively.  \\    
    $b_{ij}^{r}$, $b_{k}^{r}$ & $\mathbb{R^+}$ & The number of units of an asset $a_{ij}^{r}$ and the number of units of a forward pair $q_{k}^{r}$ bought at a recourse node $r$, respectively.  \\    
    $s_{ij}$, $s_{k}$ & $\mathbb{R^+}$ & The number of units of an asset $a_{ij}$ and the number of units of a forward pair $q_{k}$ sold in the first stage, respectively. \\ 
    $s_{ij}^{r}$, $s_{k}^{r}$ & $\mathbb{R^+}$ & The number of units of an asset $a_{ij}$ and the number of units of a forward pair $q_{k}$ sold at a recourse node $r$, respectively.  \\ 
     $x_{ij}$, $x_{k}$ & $\mathbb{B}$ & The binary decision variables for buying an asset $a_{ij}$ and a forward pair $q_{k}$ in the first stage, respectively. \\    
    $x_{ij}^{r}$, $x_{k}^{r}$ & $\mathbb{B}$ & The binary decision variables for buying an asset $a_{ij}$ and a forward pair $q_{k}$ at a recourse node $r$, respectively. \\        
    $y_{ij}$, $y_{k}$ & $\mathbb{B}$ & The binary decision variables for selling a forward pair $q_{k}$ in the first stage, respectively. \\
    $y_{ij}^{r}$, $y_{k}^{r}$ & $\mathbb{B}$ & The binary decision variables for selling a forward pair $q_{k}$ at a recourse node $r$, respectively. \\
    $z_{j}$, $z_{j}^{r}$ & $\mathbb{B}$ & The binary decision variable for investing in country $j$ in the first stage and at a recourse node $r$, respectively. \\
    \bottomrule
    \end{tabular}%
  \label{tab:notations_decvar} %
  }
\end{table}%

\subsubsection{The Model}
Here we propose the two-stage stochastic portfolio optimisation model. The proposed model comprises two stages, the first stage makes decision to buy or sell assets and forwards based on existing information  before the realisation of the uncertain asset and forward prices (equations (\ref{assetbalance}) to (\ref{binary})). Then when prices become realised, further decisions are made in the second stage to avoid constraints violations (equations (\ref{rec_assetbalance}) to (\ref{rec_binary})). A decision at this stage generally depends on a particular realisation (scenario) of the uncertain variables.

{\small
\begin{align}		
	\text{minimise}
	   \label{objectivefunction}
       & \quad \left(\alpha + \frac{1}{1-\beta}\sum\limits_{r\in N_r}p^{r}e^{r}\right) \\
    \text{subject to} \nonumber \\
    & \textit{First Stage: Portfolio Selection} \nonumber \\  
       \label{assetbalance}  
       & \quad a_{ij} = a_{ij}^0 + b_{ij}x_{ij} - s_{ij}y_{ij}, \\ 
       \label{forwardsbalance}
       & \quad q_{k} = q_{k}^0 + b_{k}x_{k} - s_{k}y_{k}, \\ 	   
       \label{mvbalance} 
       & \quad h^0 + \sum\limits_{i=1}^{A}\sum\limits_{j=1}^{C}\left(s_{ij}P_{ij}^{0}\right)y_{ij} - 
         \sum\limits_{i=1}^{A}\sum\limits_{j=1}^{C}\left(\psi_{ij}+
         \lambda_{ij}s_{ij}P_{ij}^{0}\right)y_{ij} \nonumber \\
       & \quad \phantom{{...}} + \sum\limits_{k=1}^{G}\left(s_{k}P_{k}^{0} 		 		          					 \right)y_{k} - \sum\limits_{k=1}^{G}\left(\psi_{k}+
         \lambda_{k}s_{k}P_{k}^{0}\right)y_{k} \nonumber \\  
       & \quad \phantom{{...}}= 
         \sum\limits_{i=1}^{A}\sum\limits_{j=1}^{C}\left(b_{ij}P_{ij}^{0}\right)x_{ij} -        
         \sum\limits_{i=1}^{A}\sum\limits_{j=1}^{C}\left(\pi_{ij}+
         \rho_{ij}b_{ij}P_{ij}^{0}\right)x_{ij} \nonumber \\
       & \quad \phantom{{...}} +
         \sum\limits_{k=1}^{G}\left(b_{k}P_{k}^{0}\right)x_{k} -
         \sum\limits_{k=1}^{G}\left(\pi_{k} +
         \rho_{k}b_{k}P_{k}^{0}\right)x_{k}, \\	 
       \label{mvforwards}    
       & \quad \textbf{q} = \left[q_{k} \circ P_{k}^{0}\right], \\
	   \label{forwardsmatrix}       
       & \quad \textbf{F} = \textbf{T} \circ \left(\textbf{1}^{\textrm{T}} \otimes \textbf{q}\right), \\   
       \label{cashfwd}
 	   & \quad a_{ij}^{csh} = M\sum\limits_{k=1}^{G}\left| q_{k}\right|P_k^0, \\
       \label{currencydefinition}	   
	   & \quad c_j = \sum\limits_{i=1}^{A}a_{ij}P_{ij}^0 + \sum\limits_{k=1}^{G}f_{{kj}} + a_{ij}^{csh},  \\	
	   \label{totovl}	   
	   & \quad \dfrac{1}{2}\sum\limits_{j=1}^{C}\left|\sum\limits_{k=1}^{G} f_{kj}\right| \,\leq\, 
	   V_u\sum\limits_{j=1}^{C}c_{j}, \\	
	   \label{onlybuyorsellasset}
	   & \quad x_{ij} + y_{ij} \leq 1, \\
	   \label{onlybuyorsellforwards}
	   & \quad x_{k} + y_{k} \leq 1, \\
	   \label{minbuyasset}
	   & \quad t_{ij}^{min}x_{ij} \leq b_{ij} \leq B, \\
	   \label{minsellasset}
	   & \quad t_{ij}^{min}y_{ij} \leq s_{ij} \leq B, \\
	   \label{minbuyforwards}
	   & \quad t_{k}^{min}x_{k} \leq b_{k} \leq B, \\
       \label{minsellforwards}
	   & \quad t_{k}^{min}y_{k} \leq s_{k} \leq B, \\	 	   
  	   \label{currencyexposure}  
	   & \quad c_j^{min}z_{j} \leq c_{j} \leq  c_j^{max}, \\
	   \label{assetcardinality}
 	   & \quad \sum\limits_{i=1}^{A}x_{ij} + \sum\limits_{i=1}^{A}y_{ij} \geq z_{j}; \quad \forall j \in C, \\
  	   \label{currencycardinality}
  	   & \quad \sum\limits_{j=1}^{C}z_{j} \leq K_C, \\ 	   
	   \label{forwardscardinality}
 	   & \quad \sum\limits_{k=1}^{G}x_{k} + \sum\limits_{k=1}^{G}y_{k} \leq K_G, \\ 	   
 	   \label{buysell}   	
 	   & \quad b_{ij}, b_{k}, s_{ij}, s_{k} \in \mathbb{R^+}, \\ 
 	   \label{binary}   
	   & \quad x_{ij}, x_{k}, y_{ij}, y_{k}, z_{j} \in \{0,1\}, \\ 	  
       & \textit{Second Stage: Recourse} \nonumber \\      
       \label{rec_assetbalance}  
       & \quad a_{ij}^{r} = a_{ij} + b_{ij}^{r}x_{ij}^{r} - s_{ij}^{r}y_{ij}^{r}, \\ 
       \label{rec_forwardsbalance}
       & \quad q_{k}^{r}  = q_{k} + b_{k}^{r}x_{k}^{r} - s_{k}^{r}y_{k}^{r}, \\ 
       \label{rec_mvbalance} 
       & \quad \sum\limits_{i=1}^{A}\sum\limits_{j=1}^{C}\left(s_{ij}^{r}P_{ij}^{r}\right)y_{ij}^{r} - 
         \sum\limits_{i=1}^{A}\sum\limits_{j=1}^{C}\left(\psi_{ij}+
         \lambda_{ij}s_{ij}^{r}P_{ij}^{r}\right)y_{ij}^{r} \nonumber \\
       & \quad \phantom{{...}} + \sum\limits_{k=1}^{G}\left(s_{k}^{r}P_{k}^{r} 		 		          					 \right)y_{k}^{r} - \sum\limits_{k=1}^{G}\left(\psi_{k}+
         \lambda_{k}s_{k}^{r}P_{k}^{r}\right)y_{k}^{r} \nonumber \\  
       & \quad \phantom{{...}}= 
         \sum\limits_{i=1}^{A}\sum\limits_{j=1}^{C}\left(b_{ij}^{r}P_{ij}^{r}\right)x_{ij}^{r} -        
         \sum\limits_{i=1}^{A}\sum\limits_{j=1}^{C}\left(\pi_{ij} +
         \rho_{ij}b_{ij}^{r}P_{ij}^{r}\right)x_{ij}^{r} \nonumber \\
       & \quad \phantom{{...}} +
         \sum\limits_{k=1}^{G}\left(b_{k}^{r}P_{k}^{r}\right)x_{k}^{r} -
         \sum\limits_{k=1}^{G}\left(\pi_{k} +
         \rho_{k}b_{k}^{r}P_{k}^{r}\right)x_{k}^{r}; \quad \forall r \in N_{r} \\	
       \label{rec_mvforwards}    
       & \quad \textbf{q}^{r} = \left[q_{k}^{r} \circ P_{k}^{r}\right], \\
	   \label{rec_forwardsmatrix}       
       & \quad \textbf{F}^{r} = \textbf{T} \circ \left(\textbf{1}^{\textrm{T}} \otimes \textbf{q}^{r}\right), \\   
       \label{rec_cashfwd}
 	   & \quad a_{ij}^{csh,r} = M\sum\limits_{k=1}^{G}\left| q_{k}^{r}\right|P_k^{r}, \\   
 	   \label{rec_currencydefinition}	   
	   & \quad c_{j}^{r} = \sum\limits_{i=1}^{A}a_{ij}^{r}P_{ij}^{r} + \sum\limits_{k=1}^{G}f_{kj}^{r} + a_{ij}^{csh,r},  \\	
	   \label{rec_totovl}	   
	   & \quad \dfrac{1}{2}\sum\limits_{j=1}^{C}\left|\sum\limits_{k=1}^{G} f_{kj}^{r} \right| \,\leq\, 
	   V_u\sum\limits_{j=1}^{C}c_{j}^{r}; \quad \forall r \in N_{r}, \\	
	   \label{rec_onlybuyorsellasset}
	   & \quad x_{ij}^{r} + y_{ij}^{r} \leq 1; \quad \forall r \in N_{r}, \\
	   \label{rec_onlybuyorsellforwards}
	   & \quad x_{k}^{r} + y_{k}^{r} \leq 1; \quad \forall r \in N_{r}, \\
	   \label{rec_minbuyasset}
	   & \quad t_{ij}^{min}x_{ij}^{r} \leq b_{ij}^{r} \leq B; \quad \forall r \in N_{r}, \\
	   \label{rec_minsellasset}
	   & \quad t_{ij}^{min}y_{ij}^{r} \leq s_{ij}^{r} \leq B; \quad \forall r \in N_{r}, \\
	   \label{rec_minbuyforwards}
	   & \quad t_{k}^{min}x_{k}^{r} \leq b_{k}^{r} \leq B; \quad \forall r \in N_{r}, \\
       \label{rec_minsellforwards}
	   & \quad t_{k}^{min}y_{k}^{r} \leq s_{k}^{r} \leq B; \quad \forall r \in N_{r}, \\	 	   
  	   \label{rec_currencyexposure}  
	   & \quad c_j^{min}z_{j}^{r} \leq c_{j}^{r} \leq  c_j^{max}; \quad \forall r \in N_{r}, \\
	   \label{rec_assetcardinality}
 	   & \quad \sum\limits_{i=1}^{A}x_{ij}^{r} + \sum\limits_{i=1}^{A}y_{ij}^{r} \geq z_{j}^{r}; \quad \forall j \in C,\, \forall r \in N_{r}, \\
  	   \label{rec_currencycardinality}
  	   & \quad \sum\limits_{j=1}^{C}z_{j}^{r} \leq K_C; \quad \forall r \in N_{r}, \\ 	   
	   \label{rec_forwardscardinality}
 	   & \quad \sum\limits_{k=1}^{G}x_{k}^{r} + \sum\limits_{k=1}^{G}y_{k}^{r} \leq K_G; \quad \forall r \in N_{r}, \\ 
 	   \label{rec_portwealth}
 	   & \quad W^{r} = \sum\limits_{i=1}^{A}\sum\limits_{j=1}^{C}a_{ij}^{r}P_{ij}^{r} +
 	   \sum\limits_{k=1}^{G}q_{k}^{r}P_{k}^{r} + a_{ij}^{csh,r}, \\
 	   \label{shortfall1}
 	   & \quad e^{r} \geq -\left( \frac{W^{r}}{W^{0}}-1 \right) - \alpha; \quad \forall r \in N_{r}, \\ 
 	   \label{shortfall2}
 	   & \quad e^{r} \geq 0; \quad \forall r \in N_{r}, \\
 	   \label{portreturn}
 	   & \quad \sum\limits_{r\in N_r}p^{r}\left( \frac{W^{r}}{W^{0}}-1 \right) \geq \mu, \\
 	   \label{rec_buysell}   	   
 	   & \quad b_{ij}^{r}, b_{k}^{r}, s_{ij}^{r}, s_{k}^{r} \in \mathbb{R^+}, \\
 	   \label{rec_binary}
	   & \quad x_{ij}^{r}, x_{k}^{r}, y_{ij}^{r}, y_{k}^{r}, z_{j}^{r} \in \{0,1\}. 	
\end{align}
}%
The details of the proposed optimisation model are given as follows:
\begin{itemize}

\item \textit{Objective function} - (\ref{objectivefunction}) determines an expected $\beta$-percentile CVaR of portfolio returns at the end of the second stage where $\alpha$ is the corresponding VaR value.

\item \textit{Unit balance} - the constraints (\ref{assetbalance}) and (\ref{forwardsbalance}) in the first stage and (\ref{rec_assetbalance}) and (\ref{rec_forwardsbalance}) in the second stage. The final position is calculated from initial units plus bought-in units deducted by sold-out units. 

\item \textit{Cash balance} - the constraints (\ref{mvbalance}) in the first stage and (\ref{rec_mvbalance}) in the second stage. Money for purchasing securities is strictly from initial cash plus cash received from selling securities.

\item \textit{Currency overlay construction} - the constraints (\ref{mvforwards}) and (\ref{forwardsmatrix}) in the first stage and (\ref{rec_mvforwards}) and (\ref{rec_forwardsmatrix}) in the second stage. A currency overlay is a combination of foreign exchange forwards.

\item \textit{Margin requirement of forwards} - the constraints (\ref{cashfwd}) in the first stage and (\ref{rec_cashfwd}) in the second stage. A portfolio is required to reserve some cash to retain forward positions.

\item \textit{Currency Exposure} - the constraints (\ref{currencydefinition}) in the first stage and  (\ref{rec_currencydefinition}) in the second stage. The exposure on each currency is the sum of total market values of all investments (assets and forwards) denominated in that currency.

\item \textit{Total currency overlay} - the constraints (\ref{totovl}) in the first stage and  (\ref{rec_totovl}) in the second stage. A total currency overlay of a portfolio cannot exceed a predefined limit.

\item \textit{Either buy or sell} - the constraints (\ref{onlybuyorsellasset}) and (\ref{onlybuyorsellforwards}) in the first stage and (\ref{rec_onlybuyorsellasset}) and (\ref{rec_onlybuyorsellforwards}) in the second stage. A security is not allowed to be bought or sold at the same time in order to prevent unnecessary transaction costs.

\item \textit{Bounded trading size} - the constraints (\ref{minbuyasset}), (\ref{minsellasset}), (\ref{minbuyforwards}) and (\ref{minsellforwards}) in the first stage and (\ref{rec_minbuyasset}), (\ref{rec_minsellasset}), (\ref{rec_minbuyforwards}) and (\ref{rec_minsellforwards}) in the second stage. The size of each transaction is required to be within some predefined limits.

\item \textit{Bounded curency exposure} - the constraints (\ref{currencyexposure}) in the first stage and (\ref{rec_currencyexposure}) in the second stage. Exposure of each currency is bounded to some predefined limits.

\item \textit{Cardinality constraint on number of currencies} - the constraints (\ref{assetcardinality}) and (\ref{currencycardinality}) in the first stage and (\ref{rec_assetcardinality}) and  (\ref{rec_currencycardinality}) in the second stage. The number of currencies a portfolio is exposed to cannot exceed $K_C$.

\item \textit{Cardinality constraint on number of forwards contracts} - the constraints (\ref{forwardscardinality}) in the first stage and (\ref{rec_forwardscardinality}) in the second stage. The number of forwards in the currency overlay cannot exceed $K_G$.

\item \textit{CVaR evaluation} - the constraints (\ref{rec_portwealth}), (\ref{shortfall1}) and (\ref{shortfall2}). CVaR is calculated as an expected loss beyond the VaR value $\alpha$.

\item \textit{Portfolio return target} - the constraint (\ref{portreturn}). A target return of the portfolio.

\end{itemize}

\section{Algorithm} \label{sec:algo}

The optimisation problem outlined earlier is a two-stage stochastic programming which is generally more challenging than a deterministic model due to an increasing search space from additional variables in the second stage. For example, it is proved in \cite{DyerStougie2006} that linear two-stage stochastic programming problems are $\#$P-Hard which is generally believed to be harder than the corresponding NP-complete problems.

Cardinality constraints are also included into our optimisation model which makes the problem integer-mixed and the search space discontinuous resulting in even more challenging optimisation problem \cite{StoyanKwon2011}. We initially attempted to solve our two-stage stochastic portfolio optimisation with the exact search method using CPLEX which fails to give any feasible solution within a time limit of three hours. We therefore employed a genetic algorithm to solve this problem. 

\subsection{Genetic Algorithm}
A Genetic Algorithm (GA) is a (stochastic) search algorithms originally developed by Holland \cite{Holland1975}. The idea is inspired by the basic principles of biological evolution and natural selection. GA imitates the evolution of living organisms, where the fittest individuals dominate over the weaker ones, by mimicking the biological mechanisms of evolution, such as selection, crossover and mutation.

In a GA, an initial population (usually) consists of several feasible solutions (individuals). Based on the fitness values, some individuals are probabilistically selected to remain unchanged to the next generation and some individuals are probabilistically selected to engage in the genetic operations
to produce children (offspring) for the next generation. Cross over and mutation genetic operators take part in the process of ‘‘generation production” and in successive generations the fitness values are evaluated again. This process is repeated until acceptable solutions are found or a termination criterion is satisfied. Comprehensive descriptions of GA methods can be found in \cite{Goldbergothers1989}.

Let $\Theta$ denote a population size, $\Gamma$ denote the number of generations, $\Delta$ denote a selection probability, $\Lambda$ denote a crossover rate and $\Upsilon$ denote a mutation rate. The details of the genetic algorithm applied to solve our optimisation problem is provided in Algorithm \ref{alg:GA}, and the parameter values for GA and the optimisation model are given in Section \ref{subsubsect:parameters}.

\begin{algorithm}
\caption{Genetic Algorithm for the optimisation problem.}          
\label{alg:GA} 
\begin{algorithmic}[1]
\item Initialise the zeroth generation from a randomly-weighted portfolio.
\FOR{each individual in the initial generation} 
\STATE{Evaluate the fitness values.} 
\ENDFOR
\WHILE{termination criterion is not satisfied} 
\STATE{\textbf{Elitism:} select the fittest individuals in the current generation in the current generation that are guaranteed to survive to the next generation.} 
\STATE{\textbf{Selection:} select $\Delta(\Theta-1)$ individuals to become parents for the next generation using the Stochastic Uniform Selection Method.} 
\STATE{\textbf{Crossover:} pairing the selected parents to generate $\Lambda(\Theta-1)$ children for the next generation other than the elite child using the Arithmetic Mean method.} 
\STATE{\textbf{Mutation:} $(1-\Lambda)(\Theta-1)$, or the rest of the next generation, are generated through mutation. The Adaptive Feasible Mutation is employed to add a randomly generated number to each gene (a decision variable) of the selected individuals.}
\ENDWHILE
\end{algorithmic}
\end{algorithm}

In our study, the genetic algorithm is implemented via GA Toolbox in MATLAB R2015b. A chromosome contains all the decision variables presented in Table \ref{tab:notations_decvar}. We specify the selection function as the Stochastic Uniform Selection Method. The approach lays out a line in which each parent corresponds to a section of the line of length proportional to its scaled value. The algorithm moves along the line in steps of equal size. At each step, the algorithm allocates a parent from the section it lands on. The first step is a uniform random number less than the step size.

For the crossover function, children are created from the arithmetic mean of two parents. For the mutation function, since our optimisation problem is constrained, the Adaptive Feasible Mutation is selected. The idea is to randomly generates directions (positive or negative random numbers) that are adaptive with respect to the last successful or unsuccessful generation. Other GA Toolbox configurations to solve our optimisation problem are set to the default.

\begin{figure}[H]
\begin{centering}
\includegraphics[scale=0.35]{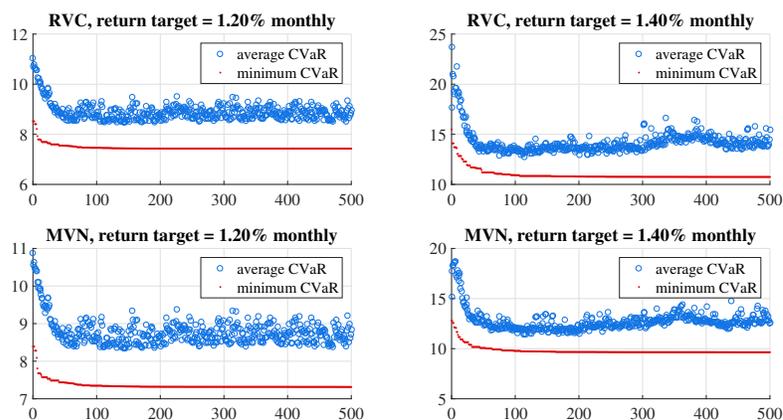}
\par\end{centering}
\caption{The evolution of fitness values (CVaR) over generations when running GA on portfolio with vine-copula-based and multivariate normal scenarios at target returns of 1.20\% and 1.40\%. The blue hollow circles represent the average of fitness values over an entire population and the red dots indicate the best fitness values (minimum CVaRs). All the optimisations converge within the maximum limit of 500 generations.}
\label{GA_500gens}
\end{figure}

Due to the probabilistic development of the solutions, GA cannot guarantee optimality even when it may be reached. To verify the reliability of solutions from GA under our settings, we examine the plot of fitness values at each generation for portfolios at target returns 1.20\% and 1.40\%. In Figure \ref{GA_500gens}, the top panel shows the results from vine-copula-based scenarios and the bottom panel displays the results from multivariate normal scenarios. The blue hollow circles represent the average of fitness values over an entire population and the red dots indicate the best fitness values (minimum CVaRs). All the optimisations (both shown and not shown in Figure \ref{GA_500gens}) converge within the maximum number of generations (500). The runtime spent on solving each portfolio is approximately 50 minutes (see more details in Table \ref{tab:runtime}). For most cases, the fitness value reaches the minimum at around the $250^{th}$ generation which gives evidence of the reliability of solutions obtained from the GA when solving our optimisation problem.

\section{Results and Discussions} \label{sec:res_disc}

\subsection{Data}
In our study, the investments of interest are in four asset classes; blue-chip stock indices, government bond indices, corporate bond indices, commodity (gold) and currencies. There are in total eighteen countries with fourteen currencies (including USD as a base currency of a portfolio). Equity investment covers all countries while government bond investments are in most countries (some are omitted due to small market sizes and lack of liquidity). For corporate bonds, only those issued in EUR, GBP and USD are selected to be invested in a portfolio. The only commodity invested in the portfolio is gold which is quoted in USD. All investments in the portfolio are specified in Table \ref{tab:asset_descriptions}.

The monthly return data are from Jan-02 to May-15. The in-sample period used for estimating scenarios covers Jan-02 to Dec-11. The out-of-sample period takes the rest of the data, from Jan-12 to May-15. Returns for government and corporate bonds are collected from MorganMarkets. Returns for currencies, stock indices and gold are collected from Bloomberg. 

\subsection{Parameter Settings} \label{subsubsect:parameters}
Key parameters for GA are set as below while the rest of GA toolbox parameters take the default values. For the optimisation model in all experiments, if not stated otherwise, the parameters are set as follows:
\begin{enumerate}[label=(\alph*)]
\item \textit{GA parameters} - The population size $\Theta = 500$, the number of generations $\Gamma = 500$, selection probability $\Delta = 10\%$ and crossover rate $\Lambda = 80\%$.
\item \textit{Model parameters} - An efficient frontier is created with 22 portfolios\footnote{When cardinality constraints are imposed, some return targets may not be achievable, the resulting efficient frontier could thus be comprised of a smaller number of portfolios.}, with the target returns $\mu = 0.55,0.60,\ldots,1.60$. The confidence level of CVaR $\beta = 95\%$. Fixed costs for buying assets and forwards $\pi_{ij} = 0.001\%$ and $\pi_{k} = 0.001\%$ of portfolio values. Fixed costs for selling assets and forwards $\psi_{ij} = 0.001\%$ and $\psi_{k} = 0.001\%$ of portfolio values. Variable costs for buying assets and selling assets denominated in USD, EUR, GBP and JPY; $\rho_{ij} = 0.01\%$, $\lambda_{ij} = 0.01\%$ while $\rho_{ij} = 0.05\%$, $\lambda_{ij} = 0.05\%$ when buying or selling assets denominated in other currencies. Variable costs for buying and selling each forward pair ($\rho_{k}$ and $\lambda_{k}$) are the historical average of bid-ask spreads collected from Bloomberg. The minimum and maximum holding positions of each asset $a_{ij}^{min} = 0.01\%$ and $a_{ij}^{max} = 100\%$, respectively. The minimum and maximum currency exposures for each currency $c_j^{min} = 0.01\%$ and $c_j^{max} = 100\%$, respectively. The minimum trading size of an asset and a forward pair $a_{ij} = 0.1\%$ and $q_k = 0.1\%$. The maximum total overlay limit $V_u = 100\%$. The cash margin requirement to hold a forward contract $M = 10\%$. At inception, a portfolio holds only cash of an initial amount $h^0 = 100,000$ US dollars. The cardinality constraint parameters on the number of currencies $K_C = 14$. The cardinality constraint parameters on the number of forwards $K_G = {\binom{14}{2}} = 91$. The number of scenarios (recourse nodes) $N_r = 1,000$.
\end{enumerate}

\subsection{Scenarios Generation Results}
In our study, we generate scenarios from two methods, RVC and MVN. The first one assumes no parametric form on a return distribution and models asset dependence structure with an R-Vine copula. The latter assumes that asset returns are normally distributed and the dependence structure is described through linear correlation. The resulting scenarios generated from the two methods are likely to be different due to different assumptions. 

We follow the Monte Carlo simulation techniques of \citet{levy2003computational} in order to construct scenarios based on a multivariate normal distribution (MVN). For generating RVC scenarios, the methodology is as described in Subsection \ref{RVC_scenario_generation}. We allow for 5 bivariate copula families including the rotated version of Clayton and Gumbel copulas in order to capture broader range of asset dependence structure. Table \ref{tab:list_bivar_copula_families} lists all bivariate copula families used in RVC scenario construction.

\begin{table}[H]
  \centering
  \caption{List of bivariate copula families and ranges of associated parameters. Student's t copula requires an additional parameter to specify its degree of freedom.}
  \resizebox{\columnwidth}{!}{
  \small{
    \begin{tabular}{lll}
    \toprule
     Copula family & Range of parameters & Range of additional parameters \\
    \midrule
     Gaussian & (-1,1) & - \\
     Student's t     & (-1,1) & (2, $\infty$) \\
     Clayton & (0, $\infty$) & - \\
     Gumbel & $[1, \infty)$ & - \\
     Frank & $\mathbb{R}\backslash\{0\}$ & - \\
     180-degree rotated Clayton copula & $(0, \infty)$ & - \\
     180-degree rotated Gumbel copula & $[1, \infty)$ & - \\
     90-degree rotated Clayton copula & $(-\infty, 0)$ & - \\
     90-degree rotated Gumbel copula & $(-\infty, -1]$ & - \\
     270-degree rotated Clayton copula & $(-\infty, 0)$ & - \\
     270-degree rotated Gumbel copula & $(-\infty, -1]$ & - \\
    \bottomrule
    \end{tabular}%
    }}
  \label{tab:list_bivar_copula_families}%
\end{table}%

In order to justify if the number of scenarios specified ($N_r = 1,000$) can capture the necessary characteristics of assets, we follow the in-sample stability inspection methodology in \citet{MitraMitraRoman2009} and \citet{KautWallace2003}. We test the in-sample stability using scenario sets of size 500, 1,000, 1,500 and 2,000. For each scenario set size, we exhibit the statistics of the 22 optimised objective values (the number of portfolios on an efficient frontier as specified in Subsection \ref{subsubsect:parameters}). Table \ref{tab:in-sample_scenarios} compares the descriptive statistics of CVaRs across different sizes of the scenario set for the two scenario generation methods. 
\begin{table}[H]
  \centering
  \small
  \caption{In-sample stability of mean-CVaR objective value for varying sizes of the scenario set. The statistics reported of each scenario set are calculated from 22 observations. The objective value is the value of minimum CVaR of a portfolio at each return target.}
    \begin{tabular}{lccccccccc}
    \toprule
          & \multicolumn{4}{c}{number of scenarios (RVC)} &       & \multicolumn{4}{c}{number of scenarios (MVN)} \\
    \cmidrule{2-5}\cmidrule{7-10}
          & 500   & 1,000  & 1,500  & 2,000  &       & 500   & 1,000  & 1,500  & 2,000 \\
    \midrule      
    average & 0.086 & 0.070 & 0.071 & 0.070 &       & 0.072 & 0.067 & 0.067 & 0.068 \\
    standard deviation & 0.056 & 0.032 & 0.032 & 0.032 &       & 0.034 & 0.027 & 0.027 & 0.027 \\
    range & 0.103 & 0.101 & 0.102 & 0.102 &       & 0.082 & 0.081 & 0.081 & 0.082 \\
    min   & 0.038 & 0.033 & 0.032 & 0.033 &       & 0.040 & 0.031 & 0.030 & 0.030 \\
    max   & 0.141 & 0.134 & 0.134 & 0.135 &       & 0.122 & 0.112 & 0.111 & 0.112 \\
    \bottomrule
    \end{tabular}%
  \label{tab:in-sample_scenarios}%
\end{table}%
It can be seen that when the number of scenarios reaches 1,000 or more, the objective values (minimum CVaRs) are relatively identical, as opposed to when the number of scenarios is 500. This is consistent with the notion of decision stability of \citet{KautWallace2003} that ``good'' scenarios should lead to stable decision. In our context, a stable decision is that portfolio allocation does not change when the number of scenarios varies. The results from Table \ref{tab:in-sample_scenarios} point out that the number of scenarios fewer than 1,000 demonstrates unstable outcomes comparing to the case of larger than 1,000 scenarios. Therefore, our earlier setting of 1,000 scenarios should convincingly provide stable portfolio decisions.

The comparison of descriptive statistics of asset returns under the scenario set of size 1,000 is given in Table \ref{tab:compare_scenarios}. Note that returns shown in the table are produced with interest-rate-adjusted to take into account the cost-of-carry of FX forwards (see equation (\ref{adjusted_return})). By comparing the minimum, mean and maximum returns of RVC and MVN scenarios to the raw data, it is noticed that assuming normal distribution on scenarios somewhat distorts original characteristics of asset returns. In constrast, in our approach where asset return distribution is not assumed and asset dependence structure is modelled by vine copula, the minimum, mean and maximum returns of raw data are relatively well preserved. The difference in generated scenarios causes different solutions for the optimisation problem.

\subsection{Experimental Studies}
This subsection aims to present the effects of different scenario generation methods and different types of cardinality constraints on risk-return profile of portfolios. All the experiments are run on Intel Core i7-5500U 3.0 GHz processor with 12.00 GB RAM under Windows 10 operating system.

\subsubsection{Efficient portfolios from Different Scenario Generation Methods}

This study highlights the effects of different assumptions made on generating scenarios. The first approach generating scenarios by assuming that return distributions of securities are normally distributed and the co-movement between assets is represented by correlations (or linear relationships). Another approach makes no assumption on distribution family, rather, it takes the empirical distribution from available historical data. The inter-relationship between assets is described in terms of dependence and is captured by copulas. The solutions obtained from two types of generated scenarios thus differ by the assumption on the shape of return distributions and linear or non-linear relationship between securities.

\begin{figure}[H]
\begin{centering}
\includegraphics[scale=0.35]{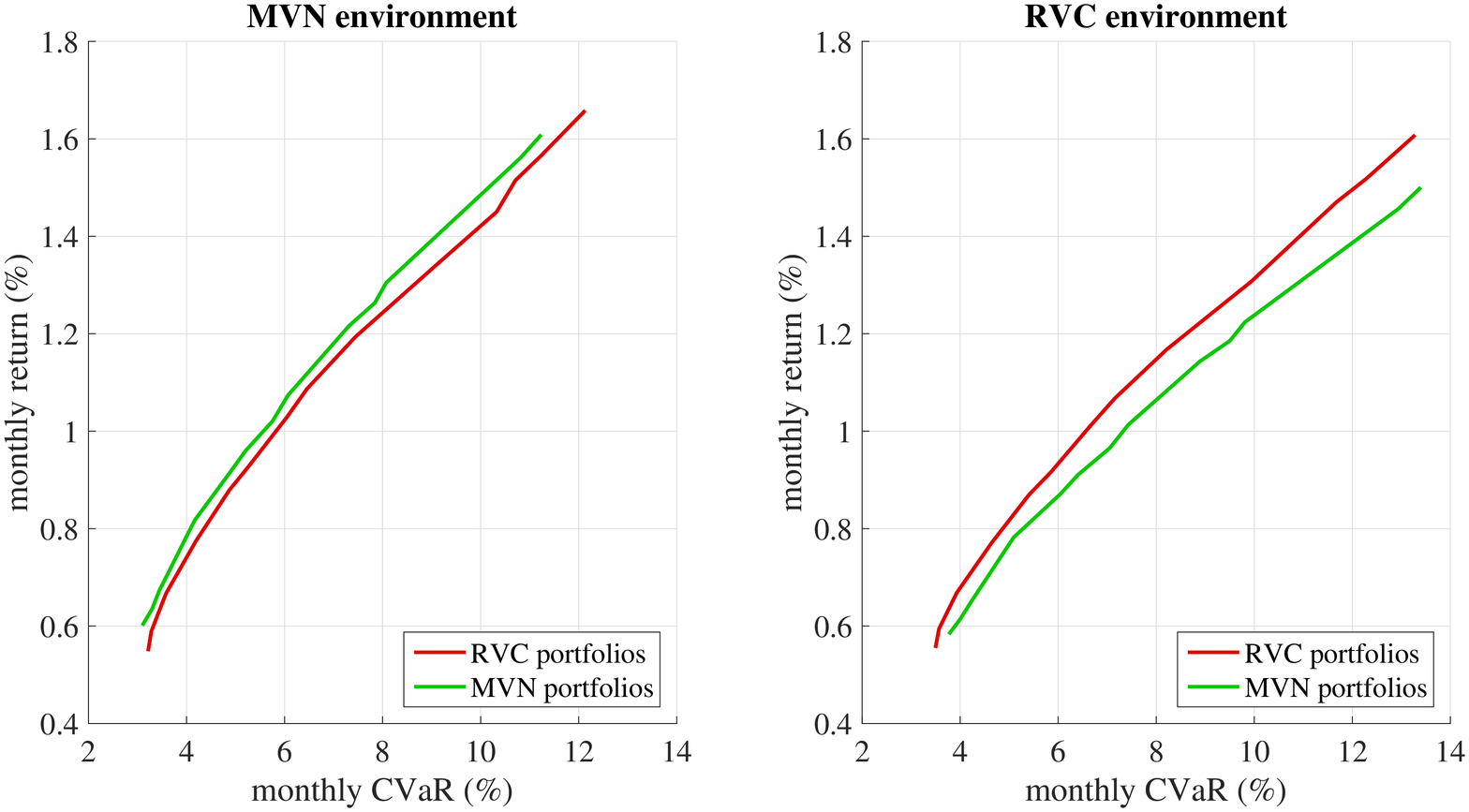}
\par\end{centering}
\caption{Comparison of efficient portfolios under different environments. The RVC portfolios are optimal under scenarios that do not assume normal distribution of returns. The MVN portfolios, on the other hand, are optimal under normal-distribution assumption. The left panel demonstrates the risk-return of efficient portfolios if returns are normally distributed and the right panel illustrates the risk-return profile of same portfolios if return distributions are not assumed Gaussian. Naturally, optimal portfolios in one environment are less efficient when evaluated in other environments.}
\label{exp1_4}
\end{figure}

Naturally, optimal portfolios are most efficient when evaluated in their own ``environment''; the one used to create them. Here we define the environment as returns generated under different underlying assumptions. For instance, return and risk of one asset can be completely different when its return distribution is assumed to be skew and fat-tailed instead of Gaussian. Hence the optimal portfolios under multivariate normal return distribution (MVN) could not be the most efficient when their risks and returns are evaluated with return scenarios generated from other methods such as regular-vine copula based scenario (RVC). This is evidently demonstrated in Figure \ref{exp1_4}.

The efficient frontiers in Figure \ref{exp1_4} is created by applying optimal allocations on returns produced from scenario generators. Naturally, optimal portfolios in one environment are less efficient when evaluated in other environments. This finding coincides with the work of Krokhmal et al. \cite{Krokhmal2002} which compare portfolios optimised under different risk measures (variance and CVaR). It is reported that CVaR optimal portfolios have higher standard deviation than that of the efficient mean-variance portfolios and the mean-variance optimal portfolios have higher CVaR than that of CVaR portfolios.

\begin{figure}[H]
\begin{centering}
\includegraphics[scale=0.35]{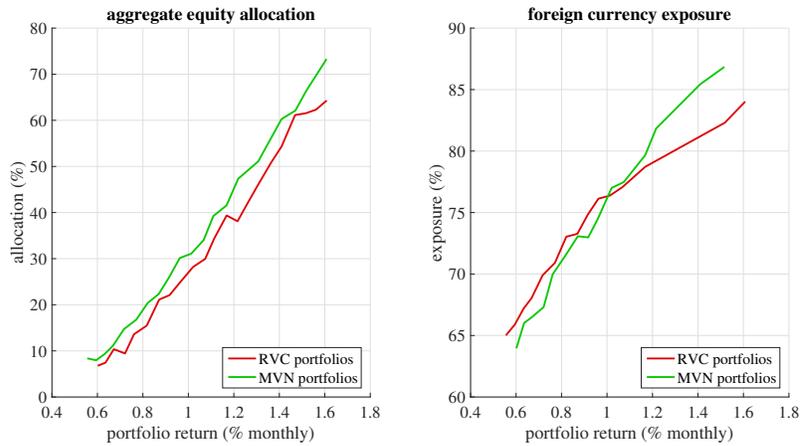}
\par\end{centering}
\caption{Equity allocation and foreign currency exposure of portfolios. The left panel shows proportion of equity (from all markets) in portfolios and the right panel presents aggregate exposure in foreign currencies (non-USD) of portfolios. It is noticed that the MVN portfolios tend to hold more equities and foreign currencies (when return is over 1\%) than the RVC counterpart.}
\label{exp1_2}
\end{figure}

Consequently, we compare optimal allocations obtained from two scenario generation methods to see if different assumptions cause deviation in optimal allocations. The left panel of Figure \ref{exp1_2} illustrates equity allocations and the right panel shows foreign currency (non-USD) exposure of portfolios. Generally, high equity holding and foreign currency exposure constitute risky portfolios. It is observed that optimal portfolios from MVN scenarios hold more equities than those optimised under RVC scenarios. For exposure on foreign currencies, although optimal portfolios from RVC scenarios more expose to foreign currencies at low-range returns, the MVN portfolios demonstrate a steep increase in foreign currency exposure when the target return is over 1\% monthly approximately.

\begin{figure}[H]
\begin{centering}
\includegraphics[scale=0.475]{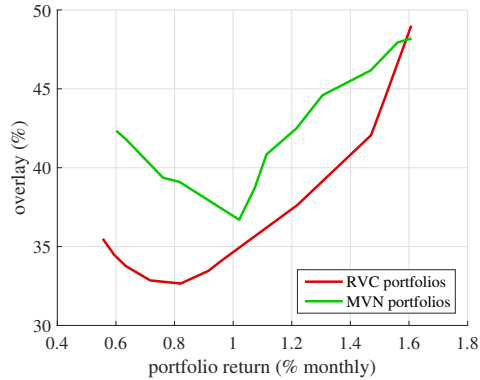}
\par\end{centering}
\caption{Currency Overlay on portfolios. The MVN portfolios demonstrate higher currency exposure adjustment though foreign exchange forwards than the RVC portfolios.}
\label{exp1_3}
\end{figure}

Foreign currency exposure is basically an aggregation of investments in foreign assets and forward positions. To investigate further on the source of foreign currency exposure, We plot the total currency overlays in Figure \ref{exp1_3}. It is shown that optimal portfolios from multivariate normal assumptions hold more forward contracts to adjust their currency exposures than optimal portfolios from copula-based scenarios for almost all return targets. The findings in Figures \ref{exp1_2} and \ref{exp1_3} conclude that portfolios optimised assuming asset returns are normally distributed tend to hold more risky investments than optimal portfolios from copula-based scenarios.

However, it should be noted that proportion of risky investments in optimal allocations cannot predict anything about portfolio performances. The reason that optimal portfolios in normal-distribution assumption hold more equities and foreign currencies could be that risks from extreme events which cannot be captured by normal distribution are overlooked and thus the exact risks could be underestimated. The proportion of risky investments is hence higher than the case that uncertainty is more-concerned. To evaluate optimal portfolios from different scenario generation methods, we perform an out-of-sample test in Section \ref{out_of_sample}.

\subsubsection{Effects of Cardinality Constraints on Risk and Return of Portfolios}
The study limits the range of currencies and the number of foreign exchange forwards to be invested in a portfolio. Although investing in multiple countries offers diversification on asset and currency exposures, investing in a small set of countries could save operational and transaction costs. Limiting the number of currencies and forwards associated in a portfolio can be implemented through cardinality constraints. In our optimisation model, the number of currencies and foreign exchange forwards in a portfolio are determined by $K_C$ and $K_G$. In this study, portfolios are optimised using RVC scenarios.

\begin{figure}[H]
\begin{centering}
\includegraphics[scale=0.35]{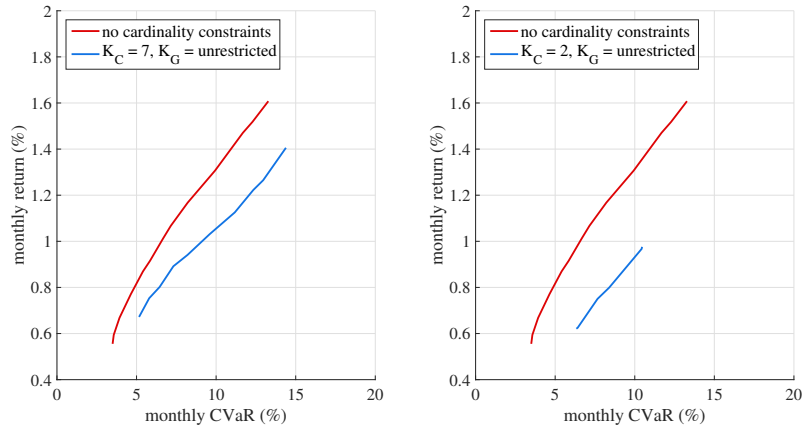}
\par\end{centering}
\caption{Comparison of portfolios without cardinality constraints and those with the cardinality constraint on the number of currencies ($K_C$). When unrestricted, $K_C = 14$ which allows portfolios to invest in all currencies. $K_C = 7$ and $K_C = 2$ then limits that portfolios can invest in 7 and 2 currencies, respectively. Although saving transaction costs, tighter restriction on the number of currencies results in lower flexibility to modify currency exposure, less efficient portfolios.}
\label{exp2_2}
\end{figure}

\begin{figure}[H]
\begin{centering}
\includegraphics[scale=0.35]{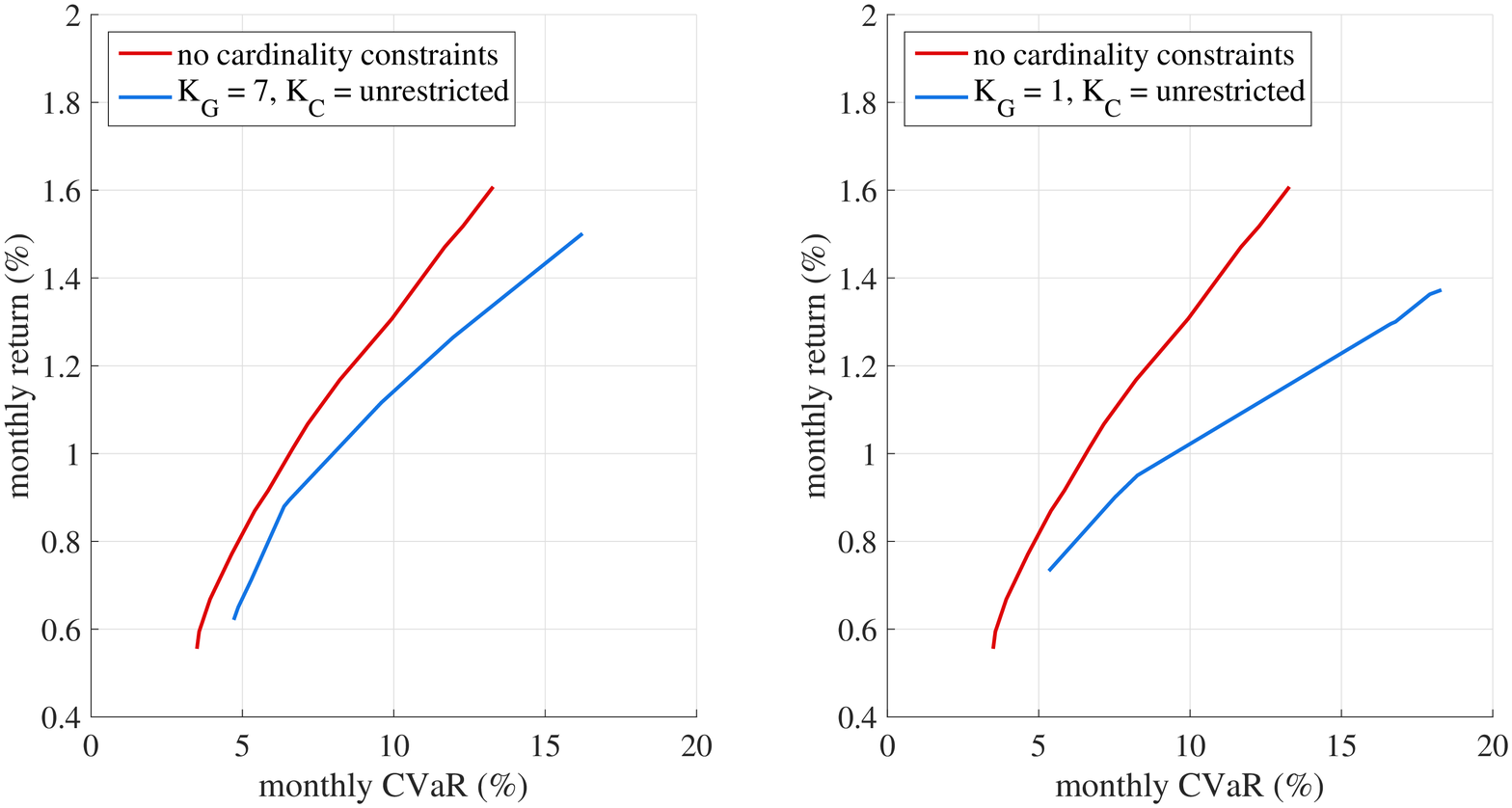}
\par\end{centering}
\caption{Comparison of portfolios without cardinality constraints and those with the cardinality constraint on the number of forwards ($K_G$). When $K_G$ is unrestricted, it implies that all forward pairs can be used to construct a currency overlay. $K_G = 7$ and $K_G = 1$ then limits that the currency overlay is a combination of 7 and 1 forward pair(s), respectively.  Although saving transaction costs, tighter restriction on the number of forwards results in lower flexibility to modify currency exposure, less efficient portfolios.}
\label{exp2_1}
\end{figure}

\begin{figure}[H]
\begin{centering}
\includegraphics[scale=0.35]{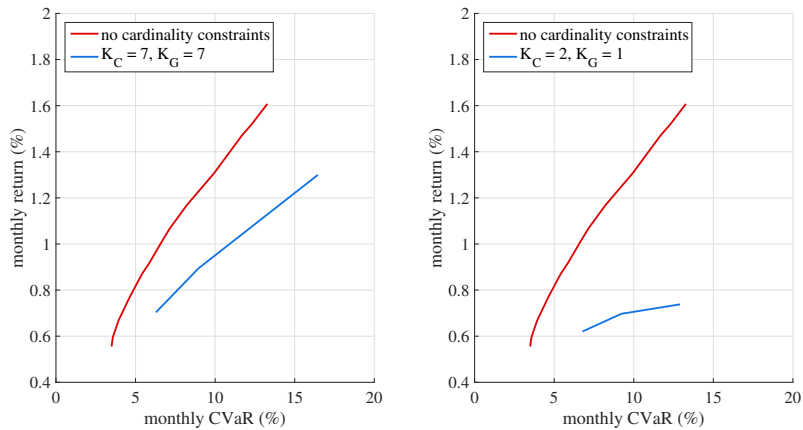}
\par\end{centering}
\caption{Comparison of portfolios without cardinality constraints and those with both the cardinality constraints on the number of currencies ($K_C$) and the number of forwards ($K_G$). $K_C = 7$ and $K_G = 7$ restrict that portfolios can invest in 7 currencies and adjust currency exposures using 7 forward pairs. $K_C = 2$ and $K_G = 1$ limit further that portfolios can invest in only 7 currencies and adjust currency exposures with only single forward pair. Although transaction costs can be reduced, tighter restriction on the number of countries and the number of forwards results in even less efficient portfolios.}
\label{exp2_3}
\end{figure}

$K_C$ indicates the number of currencies in a portfolio. Since the base currency of the portfolio is USD which is also the funding currency of forward positions, the smallest number of currencies in the portfolio is therefore two, i.e., USD and another currency. $K_G$ limits the number of foreign exchange forwards constituted in a currency overlay. Because there must be at least two or more currencies in an international portfolio, the smallest number of forwards in the portfolio is one. The plot of efficient frontiers at different levels of $K_C$ and $K_G$ are given in Figures \ref{exp2_2}, \ref{exp2_1} and \ref{exp2_3}.

The unrestricted case allows that a portfolio can invest in all currencies ($K_C = 14$) and that a currency overlay can be created from all forward pairs possible $K_G = {\binom{14}{2}} = 91$. Restricting the number of currencies that a portfolio can invest or limiting the number of forward contracts in a currency overlay make portfolios less efficient. Efficient frontiers of portfolios with cardinality constraints are truncated, some return targets are not achievable due to fewer choices of investments.

The empirical results show that imposing limits on $K_C$ causes more deviation than on $K_G$. The reason is that limiting $K_C$ affects both investment decisions in assets and currencies while limiting $K_G$ affects only choices in adjusting currency exposure. Focusing on the case that $K_G$ is restricted (Figure \ref{exp2_1}), the deviation between efficient frontiers are widened with the return targets. This is possibly due to limited choices to hedge foreign currency exposure. Basically, portfolios seek higher returns by investing in risky assets, e.g., emerging markets whose their currency returns are volatile. Hedging such exposure to local currency is an option to lower risk to portfolio but with the constraints imposed, the capacity of currency hedging is confined and hence the portfolio risk is not efficiently reduced in such cases.

Imposing both restrictions on $K_C$ and $K_G$ (Figure \ref{exp2_3}) at the same time greatly impacts risk and return of portfolios. The case that portfolios can only invest in two currencies and do a currency overlay with a single forward contract ($K_C = 2$ and $K_G = 1$) significantly limits the return ranges and flattens the efficient frontiers (implying that taking much more risk to increase the target returns). 

In terms of computational time, we report an average runtime (minutes) in running each optimal portfolio in Table \ref{tab:runtime}. Note that the sample sizes of the average runtimes are varied. The number of optimal portfolios are fewer when cardinality constraints become tighter. The cardinality constraints also result in a search space reduction which lower the solving time taken accordingly.

\begin{table}[H]
  \centering
  \small
  \caption{The average runtime (minutes) taken in solving the optimisation problem at a given level of return under different cardinality constraints configurations.  $K_C = 14$ and $K_G = 91$ indicate no restriction on the number of currencies and no restriction on the number of foreign exchange forwards, respectively. Each portfolio is run under 1,000 scenarios.}
    \begin{tabular}{lcc}    
    \toprule
    \multicolumn{1}{c}{Cardinality Constraints} & \multicolumn{1}{c}{Sample Size} &
    \multicolumn{1}{c}{Average Runtime (mins)} \\
    \midrule
       $K_C = 14$, $K_G = 91$  &  22  &  49.4 \\
       $K_C = 14$, $K_G = 7$  &  19  &  46.7 \\
       $K_C = 14$, $K_G = 1$  &  14  &  44.3 \\
       $K_C = 7$, $K_G = 91$  &  15  &  48.1 \\
       $K_C = 2$, $K_G = 91$  &  10  &  43.2 \\
       $K_C = 7$, $K_G = 7$   &  13  &  39.2 \\
       $K_C = 2$, $K_G = 1$   &  4   &  38.7 \\
    \bottomrule
    \end{tabular}%
  \label{tab:runtime}%
\end{table}%

\subsubsection{Out-of-Sample Performance} \label{out_of_sample}
This study presents cumulative returns of optimal portfolios from different scenario generation methods over the out-of-sample period (Jan-12 to May-15). Since our data frequency is monthly, there are in total 41 observations for returns and prices of all assets, currencies and forwards. The cumulative return index is constructed to compare the cumulative return by setting an initial wealth of 100 in Jan-12 then accumulating the monthly returns until May-15.

\begin{figure}[H]
\begin{centering}
\includegraphics[scale=0.35]{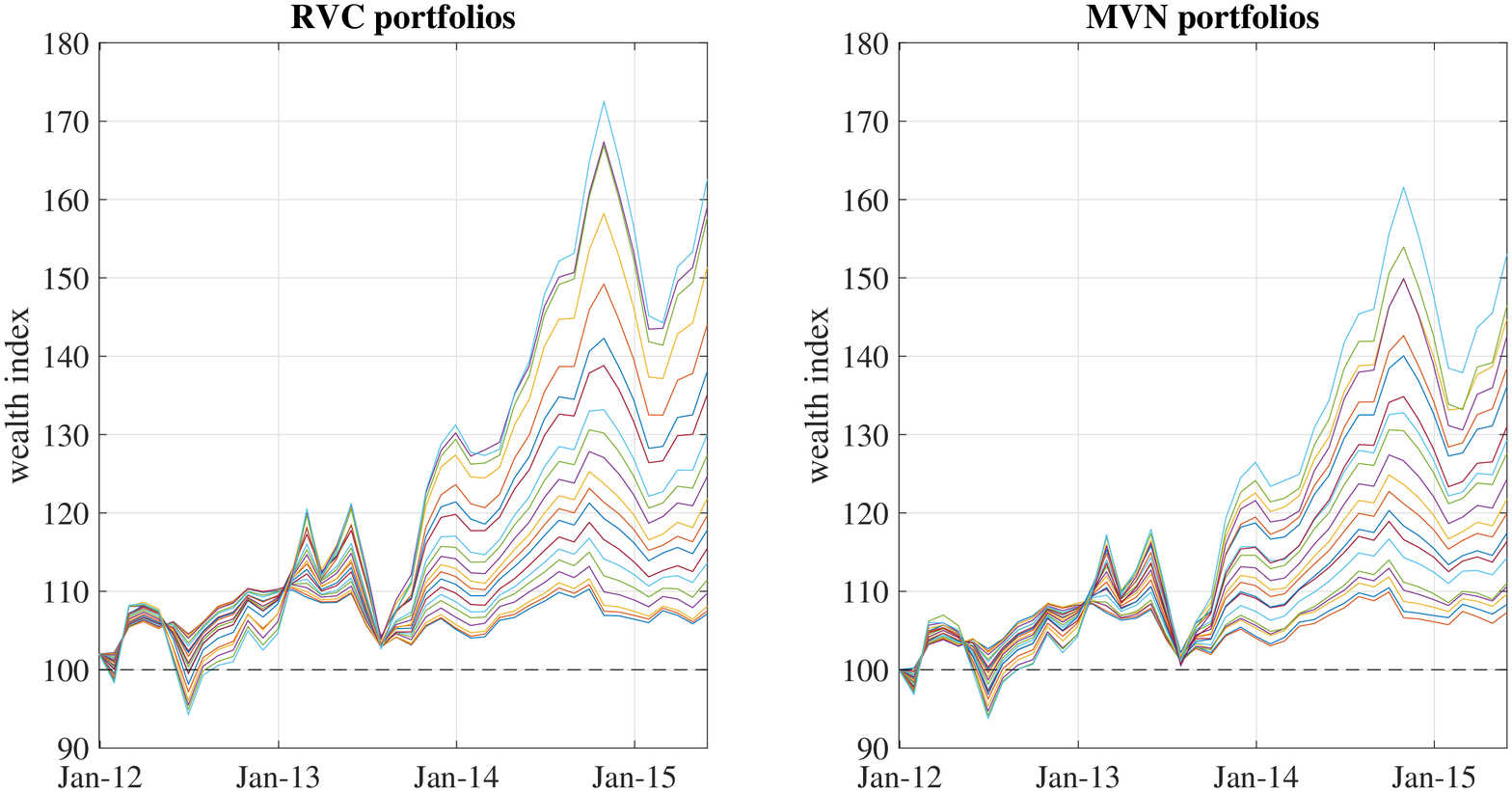}
\par\end{centering}
\caption{Cumulative wealth over the out-of-sample period (Jan-12 to May-15) of optimal portfolios from two scenario generation methods; RVC and MVN approaches. Each curve represents cumulative wealth of an optimal portfolio with respect to a given target return. Portfolios with higher risk-return profile exhibits higher final wealth with higher volatility along the period.}
\label{exp3_1}
\end{figure}

The plot of cumulative wealths is exhibited in Figure \ref{exp3_1}. Each curve represents the cumulative wealth of each optimal portfolio. The top curves are from the portfolios with high expected returns showing highly volatile paths but ending with high final wealths. The bottom curves are from the portfolios with low return targets which demonstrate more stable wealths along the period with decent wealth in May-15. It is noticed that, in the first year of the test period, portfolios with low to moderate expected return experience negative wealth while the risky portfolios do not. In addition, the gaps between portfolio wealths are tight at the beginning and become widening after late 2013. This is mainly due to stock markets rally as a results of several stimulus packages from the central banks after consecutive financial crises.

\begin{figure}[H]
\begin{centering}
\includegraphics[scale=0.35]{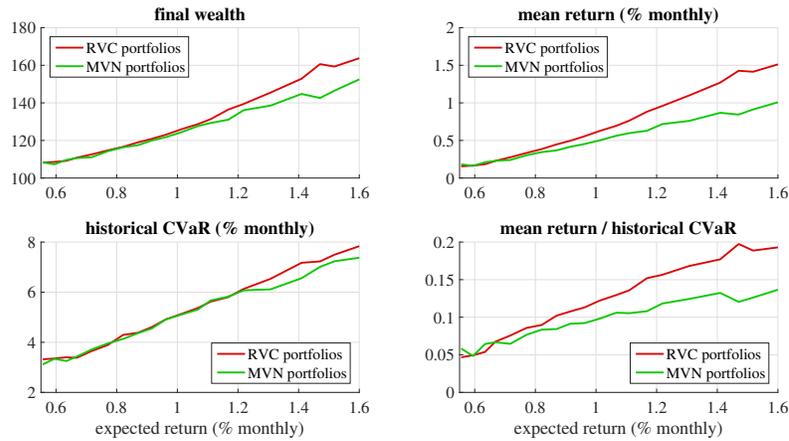}
\par\end{centering}
\caption{Performances comparison of efficient portfolios under RVC and MVN scenarios. \textit{final wealth} presents the wealth of a portfolio on May-15 given that the starting wealth is 100 on Jan-12. \textit{mean return} shows average return over the period of portfolios. \textit{historical CVaR} exhibits the average portfolio returns below $5^{th}$-percentile. \textit{mean return / historical CVaR} reports the ratio of average portfolio return over the historical CVaR.}
\label{exp3_2}
\end{figure}

To compare the performances of efficient portfolios from RVC and MVN scenarios, we present various performance evaluation measures as displayed in Figure \ref{exp3_2}. The top-left panel of Figure \ref{exp3_2} exhibits the final wealths by different levels of expected return target. It is shown that final wealths from two generation methods coincide with each other until when the target return is greater than 1\%, the RVC portfolios then shows significantly higher final wealths. The top-right panel displays the average monthly return over the test period where the copula-based portfolios demonstrate higher average return at almost all return targets.

The bottom-left panel plots the historical CVaR of portfolio returns which is calculated by averaging returns lower than the $5^{th}$-percentile of portfolio returns in the test period. It is noticed that the CVaRs over the out-of-sample period for RVC and MVN portfolios are relatively the same for low to medium return targets. Then the CVaRs of copula-based portfolios are significantly higher when portfolios reach higher returns. The plots of mean return and historical CVaR reveal that RVC portfolios show higher CVaRs but with higher returns than MVN portfolios. To see the risk-return compensation, the ratio of average return over CVaR is calculated and illustrated in the bottom-right panel. The higher value of the ratio indicates higher average return is rewarded per unit of risk taken. It is observed that RVC portfolios clearly show better risk-return reward although having higher CVaRs in general. It is only at very low return targets that MVN portfolios have better risk-return ratio due to the much lower CVaR values in the range of low returns.

\section{Conclusions} \label{sec:concl}

In this study, an optimisation model for international portfolio with currency overlay is proposed. The portfolio is integrated with a currency overlay constructed from foreign exchange forwards providing flexibility for hedging and speculating currency exposure. The portfolio structure allows asset allocation and forward positions to be optimised at the same time for portfolio optimality in terms of asset and currency exposure.

In generating scenarios for the associated stochastic optimisation problem, the key improvement is that we do not assume normality in the returns, and instead model realistic dependence structures of returns by using a regular-vine copula (RVC). The aim is to better capture the underlying risk and interdependence of securities held in a portfolio. The resulting portfolios optimised under scenarios generated from our method are compared to those obtained from a traditional approach, i.e., assuming that returns are multivariate normally distributed (MVN). The experiment results show that efficient RVC portfolios hold a lower proportion (compared to MVN solutions) of risky assets foreign currency when they are optimised using copula-based scenarios. In addition, the out-of-sample performances show that the RVC portfolios demonstrate higher risk-return reward than the MVN portfolios. The difference in risk-return compensation is not significant when portfolio return targets are under 1\% monthly but become ever more noticeable when the return targets are higher. This finding highlights the benefit of employing a regular-vine copula, particularly when portfolios hold substantial portion of risky assets, and which do not generally follow normality assumption.

\bibliographystyle{apa} 
\bibliography{Bibliography}







\appendix

\section{Sequential Method to Select an R-Vine Model} \label{apdx:SequentialMethod}
Dissmann et al. \cite{Dissmann2012} proposed that the algorithm involves searching for an appropriate R-vine tree structure, the pair-copula families, and the parameter values of the chosen pair-copula families which is summarised as follows.
\begin{enumerate}
\item Calculate the empirical Kendall’s tau for all possible variable pairs.
\item Select the tree that maximises the sum of absolute values of Kendall’s taus.
\item Select a copula for each pair and fit the corresponding parameters.
\item Transform the observations using the copula and parameters from Step 3. To obtain the transformed values.
\item Use transformed observations to calculate empirical Kendall’s taus for all possible pairs.
\item Proceed with Step 2. Repeat until the R-Vine is fully specified.
\end{enumerate}

\section{Scenario Generation using R-Vine Copula} \label{apdx:ScenarioGeneration}

\begin{enumerate}
\item \textit{Modelling marginal distributions} - The example of fitting marginal distributions to monthly returns of US treasury index, S\&P500 index and EURUSD is shown in Figure \ref{MarginalFitting}. In our scenario generation method, the empirical distribution estimates are employed to fit the return distributions. 
\par
\begin{minipage}{\linewidth}
\centering
\includegraphics[scale=0.32]{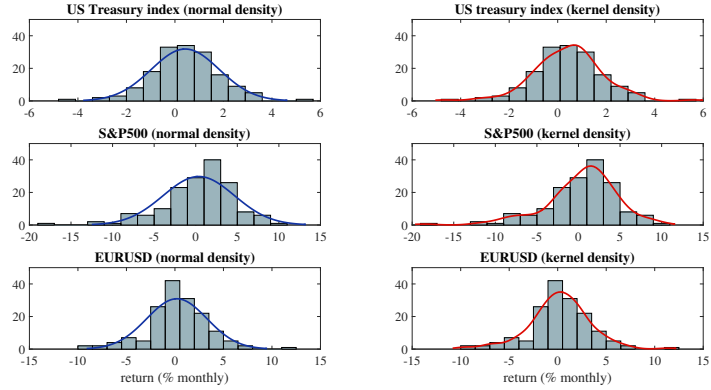}
\captionof{figure}{Marginal distribution fitting of US treasury index, S\&P500 index and EURUSD monthly returns over Jan-02 to Dec-11. The histograms illustrate distributions of raw data, the blue curves show the assumed normal distributions while the red curves demonstrate empirical distributions estimated by Epanechnikov kernel and optimal bandwidth. It can be noticed that extreme returns can be better captured by empirical distributions that normal distributions.}
\label{MarginalFitting}
\end{minipage}

\item \textit{Estimating a regular-vine copula} - The regular-vine copula specification is estimated by the sequential method of Dissmann et al. \cite{Dissmann2012} as outlined in \ref{apdx:SequentialMethod}. Let $X_1, X_2$ and $X_3$ denote the monthly returns of US treasury index, S\&P500 index and EURUSD, respectively. The estimation results from the sequential method show that the joint density decomposition is as shown in Figure \ref{SampleVine}. 
\par
\begin{minipage}{\linewidth}
\centering
\includegraphics[scale = 0.32]{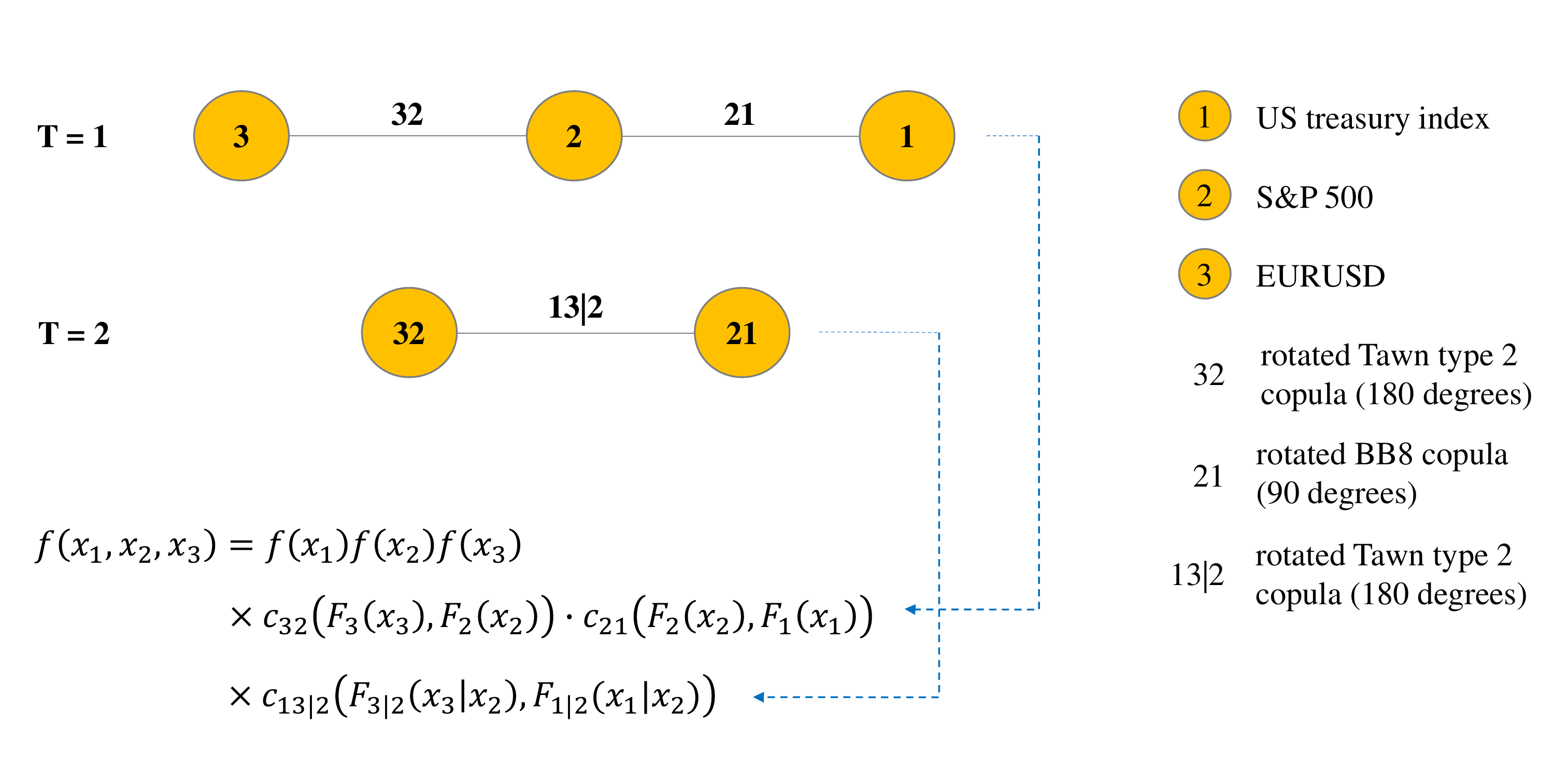}
\captionof{figure}{The regular vine copula of US government bond index, S\&P500 index and EURUSD. With three variables, there exists two trees composing the vine. The yellow nodes represent the return distributions while the edges linking the nodes represents the associated copulas. The combination of nodes and edges from all trees produces the joint density function.}
\label{SampleVine}
\end{minipage}

\item \textit{Sampling for a regular-vine density} - We simulate 1,000 scenarios for returns of each asset. Figure \ref{CopulaSim_1} display histograms of return distributions from raw data (left panels) and histograms of return distributions from simulated data (right panels). The simulated marginal distribution s exhibit similar means and variances to those from raw data. 
\par
\begin{minipage}{\linewidth}
\centering
\includegraphics[scale = 0.32]{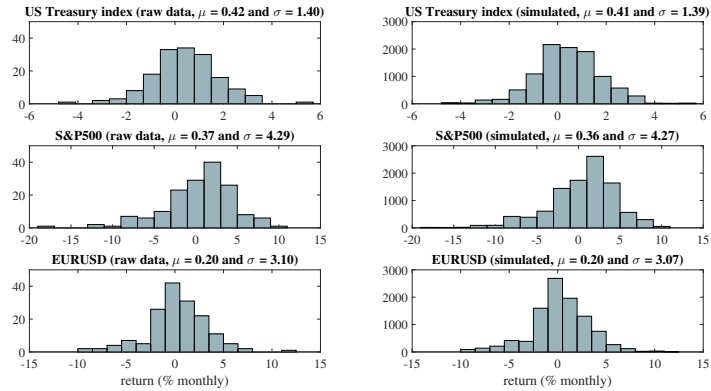}
\captionof{figure}{The return histograms of raw data (left panels) and simulated data (right panels).}
\label{CopulaSim_1}
\end{minipage}
\\

Figure \ref{CopulaSim_2} shows the bivariate return distributions of raw data and simulated data. The joint distribution shows the interdependence between asset returns which is characterised by the copula family and copula parameters estimated earlier.

\par
\begin{minipage}{\linewidth}
\centering
\includegraphics[scale=0.32]{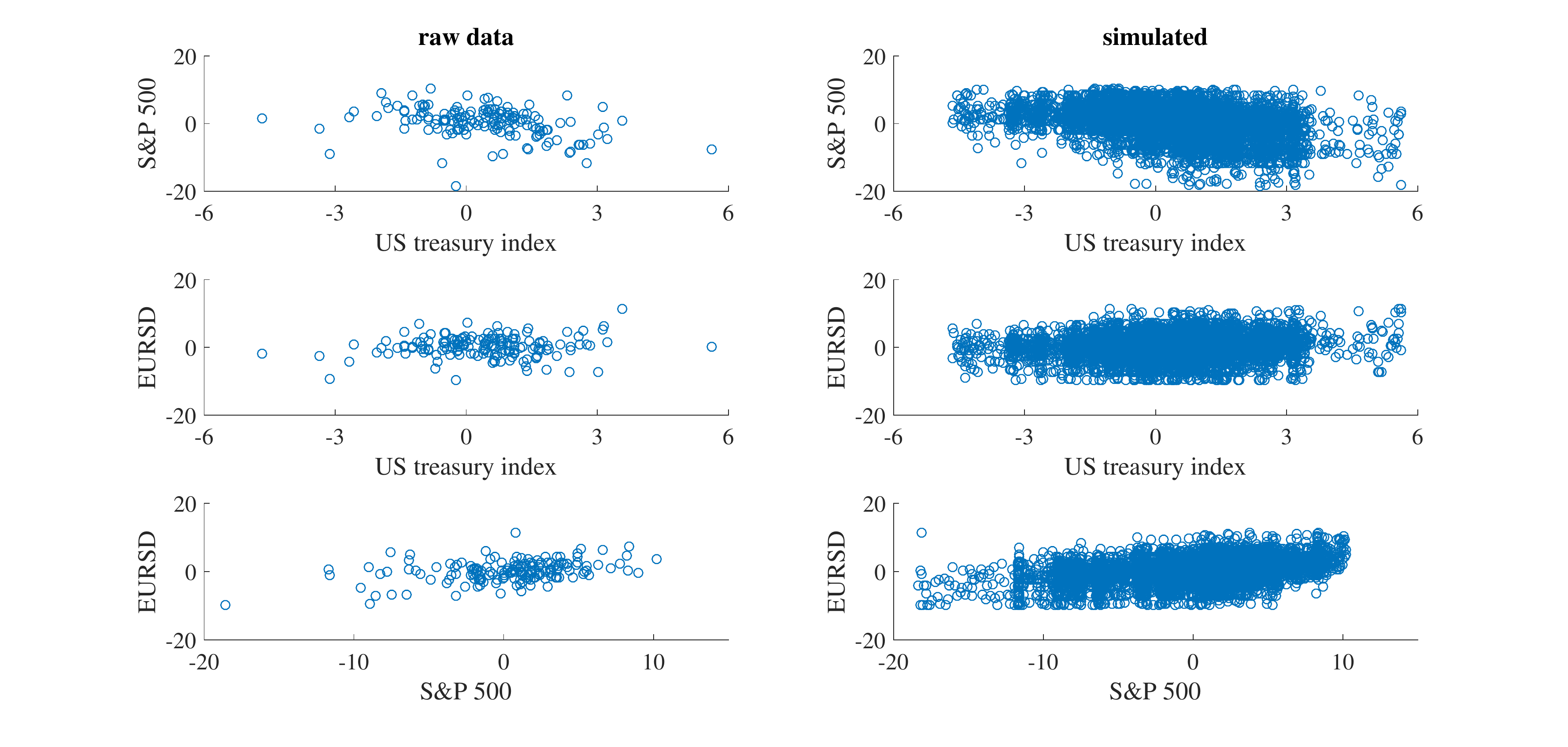} 
\captionof{figure}{\AJP{CHECK THIS USED THE CORRECT FILE} The joint return distributions by asset pairs. The left panels show the bivariate distributions of raw data and the right panels show the bivariate distributions of simulated data.}
\label{CopulaSim_2}
\end{minipage}
\end{enumerate}

\section{Asset and Currency Universe}

\begin{table}[H]
  \centering
  \caption{List of assets and currencies in a portfolio.} 
  \scriptsize{
    \begin{tabular}{lll}
    \toprule
    ID    & Notation & Description \\
    \midrule
    1     & ARS   & Argentinian Peso \\
    2     & BRL   & Brazillian Lira \\
    3     & CAD   & Canadian Dollar \\
    4     & CLP   & Chilean Peso \\
    5     & MXN   & Mexican Peso \\
    6     & EUR   & Euro \\
    7     & GBP   & Sterling Pound \\
    8     & CHF   & Swiss Franc \\
    9     & AUD   & Australian Dollar \\
    10    & INR   & Indian Rupee \\
    11    & JPY   & Japanese Yen \\
    12    & KRW   & South Korean Won \\
    13    & SGD   & Singapore Dollar \\
    14    & govt.AU & Australian government bond index (1-10 years) \\
    15    & govt.AT & Austria government bond index (1-10 years) \\
    16    & govt.BE & Belgium government bond index (1-10 years) \\
    17    & govt.CA & Canada government bond index (1-10 years) \\
    18    & govt.FR & France government bond index (1-10 years) \\
    19    & govt.DE & Germany government bond index (1-10 years) \\
    20    & govt.KR & South Korean government bond index (1-10 years) \\
    21    & govt.JP & Japan government bond index (1-10 years) \\
    22    & govt.MX & Mexico government bond index (1-10 years) \\
    23    & govt.NL & The Netherlands government bond index (1-10 years) \\
    24    & govt.SG & Singapore government bond index (1-10 years) \\
    25    & govt.UK & UK government bond index (1-10 years) \\
    26    & govt.US & US government bond index (1-10 years) \\
    27    & corp.EU & European corporate bond index (1-10 years) \\
    28    & corp.UK & UK corporate bond index (1-10 years) \\
    29    & corp.US & US corporate bond index (1-10 years) \\
    30    & stock.NL & The Netherlands stock index (AEX) \\
    31    & stock.AT & Austria stock index (ATX)  \\
    32    & stock.AU & Australia stock index top-200 (ASX200) \\
    33    & stock.BE & Belgium stock index top-20 (BEL20) \\
    34    & stock.IN & India stock index top-50 (NIFTY50) \\
    35    & stock.BR & Brazil stock index (BOVESPA) \\
    36    & stock.FR & France stock index top-40 (CAC40) \\
    37    & stock.UK & UK stock index top-100 (FTSE100) \\
    38    & stock.DE & Germany stock index (DAX) \\
    39    & stock.US & US stock index top-500 (S\&P500) \\
    40    & stock.CA & Canada stock index top-60 (S\&P/TSX 60) \\
    41    & stock.CL & Chile stock index (IPSA) \\
    42    & stock.KR & South Korea stock index (KOSPI) \\
    43    & stock.AR & Argentina stock index (MERVAL) \\
    44    & stock.MX & Mexico stock index (IPC) \\
    45    & stock.JP & Japan stock index top-225 (NIKKEI225) \\
    46    & stock.CH & Switzerland stock index (SMI) \\
    47    & stock.SG & Singapore stock index (STI) \\
    48    & XAU   & Gold (USD per troy ounce) \\
    \bottomrule
    \end{tabular}%
    }
    \label{tab:asset_descriptions}
\end{table}%

\section{Comparison of Generated Scenarios}

\begin{table}[H]
  \centering
  \caption{Descriptive statistics of raw return data, MVN return data and RVC return data. All return data are interest-rate-adjusted to take into account the cost-of-carry of FX forwards in a portfolio. The statistics for raw data are calculated on 120 observations of in-sample return data. The statistics for MVN and RVC returns are calculated on 1,000 observations from generated scenarios.}
    \tiny{
    \begin{tabular}{rrrrrrrrrrrr}
    \toprule
          & \multicolumn{3}{c}{Raw data} & \multicolumn{1}{c}{} & \multicolumn{3}{c}{MVN} & \multicolumn{1}{c}{} & \multicolumn{3}{c}{RVC} \\
    \cline{2-4} \cline{6-8} \cline{10-12}
          & min (\%) & mean (\%) & max(\%) &       & min (\%) & mean (\%) & max(\%) &       & min (\%) & mean (\%) & max(\%) \\
    \midrule       
    \multicolumn{1}{l}{USD} & 0.000 & 0.001 & 0.008 &       & 0.000 & 0.001 & 0.008 &       & 0.000 & 0.001 & 0.008 \\
    \multicolumn{1}{l}{ARS} & -49.38 & -1.17 & 7.71  &       & -21.96 & -1.13 & 19.60 &       & -49.38 & -1.05 & 7.70 \\
    \multicolumn{1}{l}{BRL} & -20.05 & -0.03 & 15.18 &       & -19.66 & 0.01  & 17.15 &       & -20.01 & -0.04 & 15.18 \\
    \multicolumn{1}{l}{CAD} & -12.56 & 0.21  & 8.42  &       & -12.51 & 0.24  & 10.33 &       & -12.43 & 0.20  & 8.40 \\
    \multicolumn{1}{l}{CLP} & -18.08 & 0.11  & 7.10  &       & -13.64 & 0.15  & 12.52 &       & -18.08 & 0.16  & 7.09 \\
    \multicolumn{1}{l}{MXN} & -14.66 & -0.28 & 7.86  &       & -10.86 & -0.22 & 11.52 &       & -14.66 & -0.24 & 7.84 \\
    \multicolumn{1}{l}{EUR} & -9.73 & 0.20  & 11.42 &       & -11.08 & 0.22  & 13.07 &       & -9.72 & 0.20  & 11.42 \\
    \multicolumn{1}{l}{GBP} & -9.35 & 0.07  & 8.78  &       & -11.24 & 0.06  & 9.71  &       & -9.34 & 0.05  & 8.78 \\
    \multicolumn{1}{l}{CHF} & -11.03 & 0.41  & 15.56 &       & -12.29 & 0.38  & 12.14 &       & -11.02 & 0.40  & 15.56 \\
    \multicolumn{1}{l}{AUD} & -16.36 & 0.34  & 9.20  &       & -14.73 & 0.38  & 13.88 &       & -16.33 & 0.35  & 9.20 \\
    \multicolumn{1}{l}{INR} & -8.07 & -0.14 & 7.40  &       & -9.03 & -0.13 & 8.25  &       & -8.06 & -0.11 & 7.40 \\
    \multicolumn{1}{l}{JPY} & -8.22 & 0.09  & 7.93  &       & -10.68 & 0.04  & 11.27 &       & -8.22 & 0.10  & 7.91 \\
    \multicolumn{1}{l}{KRW} & -12.12 & 0.18  & 16.63 &       & -11.51 & 0.16  & 12.22 &       & -12.12 & 0.21  & 16.63 \\
    \multicolumn{1}{l}{SGD} & -7.22 & 0.22  & 4.56  &       & -5.61 & 0.23  & 7.78  &       & -7.20 & 0.25  & 4.56 \\
    \multicolumn{1}{l}{govt.AU} & -2.47 & 0.54  & 4.12  &       & -3.74 & 0.51  & 4.45  &       & -2.46 & 0.54  & 4.12 \\
    \multicolumn{1}{l}{govt.AT} & -2.35 & 0.51  & 4.58  &       & -4.13 & 0.49  & 6.20  &       & -2.35 & 0.49  & 4.58 \\
    \multicolumn{1}{l}{govt.BE} & -3.74 & 0.52  & 6.83  &       & -4.61 & 0.51  & 6.21  &       & -3.73 & 0.50  & 6.83 \\
    \multicolumn{1}{l}{govt.CA} & -2.09 & 0.48  & 4.72  &       & -4.22 & 0.46  & 5.96  &       & -2.09 & 0.46  & 4.72 \\
    \multicolumn{1}{l}{govt.FR} & -2.65 & 0.48  & 4.41  &       & -4.21 & 0.46  & 5.55  &       & -2.64 & 0.46  & 4.41 \\
    \multicolumn{1}{l}{govt.DE} & -1.89 & 0.47  & 3.95  &       & -4.16 & 0.44  & 5.72  &       & -1.88 & 0.45  & 3.95 \\
    \multicolumn{1}{l}{govt.KR} & -2.98 & 0.52  & 6.82  &       & -3.45 & 0.50  & 5.07  &       & -2.97 & 0.50  & 6.82 \\
    \multicolumn{1}{l}{govt.JP} & -2.22 & 0.16  & 1.89  &       & -2.01 & 0.16  & 2.17  &       & -2.22 & 0.15  & 1.89 \\
    \multicolumn{1}{l}{govt.MX} & -5.31 & 0.78  & 6.26  &       & -7.02 & 0.77  & 7.09  &       & -5.31 & 0.77  & 6.26 \\
    \multicolumn{1}{l}{govt.NL} & -2.15 & 0.49  & 4.48  &       & -4.22 & 0.46  & 5.99  &       & -2.15 & 0.47  & 4.48 \\
    \multicolumn{1}{l}{govt.SG} & -3.15 & 0.29  & 3.71  &       & -3.38 & 0.26  & 4.44  &       & -3.15 & 0.27  & 3.71 \\
    \multicolumn{1}{l}{govt.UK} & -4.71 & 0.52  & 5.48  &       & -6.13 & 0.48  & 7.25  &       & -4.71 & 0.51  & 5.48 \\
    \multicolumn{1}{l}{govt.US} & -4.68 & 0.42  & 5.63  &       & -5.18 & 0.40  & 5.82  &       & -4.66 & 0.41  & 5.63 \\
    \multicolumn{1}{l}{corp.EU} & -2.78 & 0.40  & 2.69  &       & -3.10 & 0.39  & 3.80  &       & -2.75 & 0.39  & 2.69 \\
    \multicolumn{1}{l}{corp.UK} & -4.96 & 0.52  & 7.95  &       & -5.60 & 0.48  & 7.03  &       & -4.96 & 0.50  & 7.95 \\
    \multicolumn{1}{l}{corp.US} & -8.06 & 0.03  & 8.45  &       & -7.03 & 0.01  & 7.13  &       & -7.99 & 0.01  & 8.45 \\
    \multicolumn{1}{l}{stock.NL} & -22.62 & -0.02 & 14.57 &       & -23.19 & 0.09  & 23.86 &       & -22.60 & 0.04  & 14.54 \\
    \multicolumn{1}{l}{stock.AT} & -32.59 & 0.51  & 13.55 &       & -25.35 & 0.65  & 23.33 &       & -32.41 & 0.60  & 13.54 \\
    \multicolumn{1}{l}{stock.AU} & -13.54 & 0.33  & 7.06  &       & -13.47 & 0.40  & 13.83 &       & -13.52 & 0.37  & 7.05 \\
    \multicolumn{1}{l}{stock.BE} & -24.09 & 0.17  & 13.51 &       & -18.61 & 0.25  & 19.19 &       & -23.93 & 0.24  & 13.48 \\
    \multicolumn{1}{l}{stock.IN} & -27.30 & 1.32  & 24.89 &       & -23.67 & 1.43  & 29.73 &       & -27.05 & 1.40  & 24.86 \\
    \multicolumn{1}{l}{stock.BR} & -28.50 & 0.89  & 16.48 &       & -29.83 & 1.03  & 30.20 &       & -28.33 & 0.93  & 16.44 \\
    \multicolumn{1}{l}{stock.FR} & -19.23 & 0.05  & 12.59 &       & -19.67 & 0.14  & 19.78 &       & -19.00 & 0.10  & 12.57 \\
    \multicolumn{1}{l}{stock.UK} & -13.95 & 0.18  & 8.30  &       & -15.16 & 0.28  & 16.50 &       & -13.93 & 0.24  & 8.30 \\
    \multicolumn{1}{l}{stock.DE} & -29.33 & 0.50  & 19.37 &       & -24.20 & 0.62  & 23.21 &       & -29.09 & 0.58  & 19.33 \\
    \multicolumn{1}{l}{stock.US} & -18.56 & 0.37  & 10.23 &       & -18.29 & 0.48  & 15.22 &       & -18.35 & 0.43  & 10.20 \\
    \multicolumn{1}{l}{stock.CA} & -18.55 & 0.43  & 10.62 &       & -18.55 & 0.54  & 15.21 &       & -18.48 & 0.46  & 10.45 \\
    \multicolumn{1}{l}{stock.CL} & -10.07 & 0.77  & 14.92 &       & -15.16 & 0.79  & 19.71 &       & -10.07 & 0.77  & 14.91 \\
    \multicolumn{1}{l}{stock.KR} & -26.31 & 0.70  & 12.68 &       & -21.92 & 0.77  & 22.31 &       & -26.26 & 0.83  & 12.68 \\
    \multicolumn{1}{l}{stock.AR} & -45.81 & 2.32  & 39.67 &       & -33.43 & 2.48  & 38.24 &       & -45.70 & 2.38  & 39.64 \\
    \multicolumn{1}{l}{stock.MX} & -19.67 & 1.22  & 12.38 &       & -27.23 & 1.35  & 20.52 &       & -19.54 & 1.27  & 12.33 \\
    \multicolumn{1}{l}{stock.JP} & -27.22 & 0.39  & 12.09 &       & -23.55 & 0.52  & 23.67 &       & -26.98 & 0.52  & 12.09 \\
    \multicolumn{1}{l}{stock.CH} & -14.03 & 0.22  & 10.61 &       & -16.16 & 0.31  & 16.77 &       & -13.97 & 0.28  & 10.58 \\
    \multicolumn{1}{l}{stock.SG} & -27.36 & 0.48  & 19.30 &       & -17.75 & 0.60  & 18.84 &       & -27.29 & 0.58  & 19.27 \\
    \multicolumn{1}{l}{XAU} & -16.27 & 0.91  & 11.77 &       & -16.98 & 0.94  & 19.10 &       & -14.97 & 0.90  & 11.45 \\
    \bottomrule
    \end{tabular}%
    }
  \label{tab:compare_scenarios}%
\end{table}%

\end{document}